\documentclass{article}

\title{An Investigation into  Distance Measures in Cluster Analysis}
\author{Zoe Shapcott }
\date{April 2024}

\usepackage[margin=2cm, top=2cm, bottom=2cm]{geometry}
\usepackage[tbtags]{amsmath}
\usepackage{amssymb}
\usepackage{amsthm}
\usepackage{graphicx}
\usepackage{listings}
\usepackage{xcolor} 
\usepackage[colorlinks = true,
            linkcolor = blue,
            urlcolor  = blue,
            citecolor = blue,
            anchorcolor = blue,
            breaklinks = true]{hyperref}
\usepackage[sorting=none]{biblatex}
\usepackage{caption}
\usepackage{subcaption}
\usepackage{adjustbox}
\usepackage{physics}

\numberwithin{figure}{section}
\numberwithin{equation}{section}

\usepackage[many]{tcolorbox} 
\definecolor{main}{HTML}{000000} 
\definecolor{sub}{HTML}{ffffff}  
\definecolor{border}{HTML}{0000ff} 
\definecolor{codegreen}{rgb}{0,0.6,0}

\definecolor{boxBborder}{HTML}{d8d8d8}
\newtcolorbox{BoxA}{
    fontupper = \color{main}, 
    boxrule = 1.5pt,
    colframe = border,
    colback = sub, 
    rounded corners,
    arc = 5pt 
}
\newtcolorbox{BoxB}{
    fontupper = \color{main}, 
    boxrule = 1.5pt,
    colframe = boxBborder,
    colback = sub, 
    rounded corners,
    arc = 5pt 
}

\begin{document}

\maketitle

\begin{abstract}
    This report provides an exploration of different distance measures that can be used with the \(K\)-means algorithm for cluster analysis. Specifically, we investigate the Mahalanobis distance, and critically assess any benefits it may have over the more traditional measures of the Euclidean, Manhattan and Maximum distances. We perform this by first defining the metrics, before considering their advantages and drawbacks as discussed in literature regarding this area.  We apply these distances, first to some simulated data and then to subsets of the \emph{Dry Bean} dataset \cite{misc_dry_bean_dataset_602}, to explore if there is a better quality detectable for one metric over the others in these cases. One of the sections is devoted to analysing the information obtained from ChatGPT in response to prompts relating to this topic.
\end{abstract}

\section{Introduction}\label{sec:Introduction}
A popular method for cluster analysis is the \(K\)-means algorithm. As part of this process we require a metric that can be used to evaluate the relative `closeness' of data points. In this report, we explore a few different distance measures that can be used for this purpose. We start by introducing some general notation that will be used throughout (Section \ref{subsec:Notation}), before providing definitions of all the distance measures that we will consider (Section \ref{subsec:DistanceMeasures}). Then, in Section \ref{subsec:KMeansClustering}, we give a brief outline of the \(K\)-means algorithm and some of the techniques we use in our applications of the distance measures. 

Following this, in Section \ref{sec:CriticalAssessment}, we provide a critical assessment of some of the literature discussing and applying different distance measures for cluster analysis. Specifically, we aim to explore the Mahalanobis distance and how frequently or successfully it is implemented compared to other distance measures. Furthermore, we utilise ChatGPT and obtain its suggestions in relation to prompts regarding this topic. This will enable us to gain an appreciation of the scope and accuracy of its knowledge in this area. 

Finally, Section \ref{sec:Applications} demonstrates some applications of these distance measures for cluster analysis, by performing \(K\)-means clustering on both a simulated and real-life dataset. We investigate the results gained qualitatively through the use of scatter graphs and quantitatively by analysing the accuracy of the clustering. We can achieve this, since for our synthetic examples we already know the desired clustering beforehand. Note that the programming language R is used throughout this section, and the specific code used can be seen in Appendix \ref{sec:Appendix}.

\subsection{Notation}\label{subsec:Notation}
Throughout this report, fix \(d > 0\) and integer as the dimension of the space we are considering. Additionally, note that we denote a \(d\)-dimensional vector \(\mathbf{x}\) by 
\begin{equation}
    \mathbf{x} = (x_{1},\dotsc,x_{d})^{T}.
\end{equation}
Furthermore, given a set of \(n\) points \(X = \{\mathbf{x}_{1},\dotsc,\mathbf{x}_{n}\}\), where \(\mathbf{x}_{i} \in \mathbb{R}^{d}\) for \(i=1,\dotsc,n\), we denote the sample mean by
\begin{equation}
    \mu = (\mu_{1},\dotsc,\mu_{d})^{T}.
\end{equation}
We also denote the sample covariance matrix by \(W\). The usual inverse of \(W\) when \(W\) is non-singular is then represented as \(W^{-1}\). There are modifications that can be applied to the forthcoming definitions when \(W\) is singular, or close to singular, as can often happen in high-dimensional settings \cite{gillard2022simplicial,gillard2023polynomial}. This will be mentioned further in Sections \ref{subsec:DistanceMeasures} and \ref{sec:CriticalAssessment}. We now progress to defining the distance measures between points in \(\mathbb{R}^{d}\) that we consider throughout the remainder of this report.

\subsection{Distance Measures}\label{subsec:DistanceMeasures}
The theory of cluster analysis depends on us being able to measure the relative `closeness' between two points in \(d\)-dimensional space. This is because the broad aim of clustering is to try to group similar data points together. That is, we would like to partition the data into \(K\) groups such that the intracluster distances (i.e the distances between points in the same group) are small, but the intercluster distances (i.e the distances between points in different groups) are much larger by comparison. Hence, we can see why the metric used to measure the distance between points may be an influential factor in the performance of clustering algorithms. Note that \(K \in \mathbb{N}\) is a user-defined parameter. Thus, care should be taken to ensure that the value of \(K\) is chosen appropriately, since this will also impact the results.

There are many different methods of defining the distance between points in \(\mathbb{R}^{d}\), where some are more typical than others. We remark that there are also different distance measures that can be used when considering different types of data. For example, there are various methods for defining the distance between binary or categorical data points, such as matching coefficient measures \cite[page 47]{EverittClusterAnalysis}, Here, however, we focus only on the case of continuous numerical variables, and so just define appropriate distance measures for this scenario accordingly. 

The first distance measure we give is the very famous and commonly implemented Euclidean distance. Consider two points \(\mathbf{x} = (x_{1},\dotsc,x_{n})^{T}\) and \(\mathbf{y}=(y_{1},\dotsc,y_{n})^{T} \in \mathbb{R}^{d}\). The square of the Euclidean distance between \(\mathbf{x}\) and \(\mathbf{y}\) is then defined as 
\begin{equation}
    d_{E}^{2} = \sum_{i=1}^{d}(x_{i}-y_{i})^{2},
    \label{eq:Euclidean}
\end{equation}
or, equivalently, as
\begin{equation}
    d_{E}^{2} = (\mathbf{x}-\mathbf{y})^{T}(\mathbf{x}-\mathbf{y}),
    \label{eq:EuclideanVec}
\end{equation}
(see, for example, \cite{gillard2022simplicial}, \cite[pages 49 - 50]{EverittClusterAnalysis}).

The second distance measure we consider is the similarly defined Manhattan distance, which is based on the \(L_{1}\)-norm \cite[page 50]{EverittClusterAnalysis}, \cite[page 26]{kassambara2017practical}. The definition of the Manhattan distance is 
\begin{equation}
    d_{\text{Manhattan}} = \sum_{i=1}^{d} \abs{x_{i} - y_{i}}.
    \label{eq:Manhattan}
\end{equation}

We may also note that both the Euclidean and Manhattan distance measures are in fact special cases of the broader class of Minkowski distance measures, which are parameterised by \(p \ge 1\). The definition is
\begin{equation}
    d_{\text{Minkowski}} = \left(\sum_{i=1}^{d} \abs{x_{i} - y_{i}}^{p}\right)^{1/p}.
    \label{eq:Minkowski}
\end{equation}
Thus, we can see that the Euclidean distance is equivalent to the Minkowski distance with parameter \(p = 2\), whilst the Manhattan distance corresponds to the Minkowski distance with \(p = 1\) \cite[page 50]{EverittClusterAnalysis}.

Letting \(p \to \infty\) in the above equation, we obtain the third distance measure we consider. This is known as the Chebyshev or Maximum distance. It is defined as 
\begin{equation}
    d_{\text{max}} = \max_{i=1,\dotsc,d}\abs{x_{i} - y_{i}},
    \label{eq:Maximum}
\end{equation}
see, for example, \cite{WikiChebyshevDistance}.

Now, we can generalise all of the above definitions given for two points in \(\mathbb{R}^{d}\) to consider the distance between a point \(\mathbf{x} \in \mathbb{R}^{d}\) and a set \(X\) of \(n\) points in \(\mathbb{R}^{d}\) (as defined in Section \ref{subsec:Notation}). To do this, we simply replace \(\mathbf{y}\) in equation (\ref{eq:EuclideanVec}) and \(y_{i}\) in equations (\ref{eq:Euclidean}), (\ref{eq:Manhattan}), (\ref{eq:Minkowski}) and (\ref{eq:Maximum}) by \(\mu\) and \(\mu_{i}\) respectively, where \(\mu\) is the sample mean of \(X\) \cite{gillard2022simplicial}.

The final distance measure we consider is the Mahalanobis distance. A problem with all of the above defined distance measures is that they do not take into account any correlations that may be present between the variables, instead simply assuming that they are spherically distributed. This may be unrealistic in many scenarios, and the Mahalanobis distance is one example of a distance measure that tries to overcome this issue. The squared Mahalanobis distance between a point \(\mathbf{x}\) and a set \(X\) with mean \(\mu\) and covariance matrix \(W\) is defined as
\begin{equation}
    d_{M}^{2} = (\mathbf{x} - \mu)^{T} W^{-1} (\mathbf{x} - \mu),
    \label{eq:Mahalanobis}
\end{equation}
see, for example, \cite{gillard2022simplicial}. It can also be seen by comparing between equations (\ref{eq:EuclideanVec}) and (\ref{eq:Mahalanobis}), that the Mahalanobis and Euclidean distances are equivalent in the specific case that there are no correlations present in the data and the variances are all equal to one. That is when \(W = I_{d}\), where \(I_{d}\) is the \(d\)-dimensional identity matrix, the Euclidean and Mahalanobis distances are the same.

We remark that the necessity of inverting the sample covariance matrix can present issues if \(W\) is singular or close to singular, which can often be the case when working in high dimensions. For this scenario, the Mahalanobis distance can be generalised by using for example the Moore-Penrose pseudo-inverse of the sample covariance matrix instead, giving what is known as the pseudo Mahalanobis distance \cite{gillard2022simplicial}.

For a final note on distance measures for continuous numerical data, we mention that there are other classes of distance measures that can be used. For example, there are the \(k\)-simplicial distances defined in \cite{pronzato2017extended,PRONZATO2018276}, or the \(k\)-minimal-variance distances of \cite{gillard2022simplicial}, where in both cases, \(k\) is a user-defined parameter. Additionally, it is also possible to use correlation metrics such as Pearson or Spearman correlation distances (see, for example, \cite[page 26]{kassambara2017practical}).

The distance measure chosen to be used as part of a clustering algorithm can have dramatic effects on the results obtained. In Sections \ref{sec:CriticalAssessment} and \ref{sec:Applications} we explore literature and applications respectively to investigate this impact. Before this, however, we briefly outline the clustering method we utilise in Section \ref{sec:Applications}. 

\subsection{\texorpdfstring{$K$}{K}-Means Clustering Algorithm}\label{subsec:KMeansClustering}
There are several different ways to perform cluster analysis, for example partitioning clustering procedures or hierarchical clustering ones. For our explorations, we use the method of \(K\)-means clustering. To apply this, it is important to first make sure that the data is clean with no missing entries. Additionally, since this method can be sensitive to the scale of the data, it is also important to standardise the data first so that the units of the variable do not impact the results \cite{kassambara2017practical}. Note that this standardisation does not remove any correlations present in the dataset. 

For the case of the Euclidean, Manhattan or Maximum distance measures, the main steps of the general algorithm as implemented in R, can be described as follows \cite{kassambara2017practical}:
\begin{enumerate}
    \item Specify the number of clusters \(K\), the maximum number of iterations and the number of starts.
    \item Randomly choose \(K\) points from the dataset to be the initial centroids (cluster centres).
    \item Allocate each data point to its nearest cluster based on choosing the minimum distance between it and all cluster centroids, using the chosen distance measure.
    \item Calculate new cluster centroids based on the points assigned to each cluster in this iteration.
    \item Repeat the above two steps up to a number of times equal to the maximum number of iterations set at the beginning.
    \item Repeat from step (2) a number of times equal to that set at the beginning. 
\end{enumerate}

For the Mahalanobis distance, a slightly different method is employed, given that we need some sort of initial assignment to the clusters to calculate the covariance matrices. We follow the method given in \cite{gillard2022simplicial}, which can be described as follows. 
\begin{enumerate}
    \item Specify the number of clusters \(K\) and the number of iterations of the Mahalanobis stage to be performed.
    \item Carry out \(K\)-means clustering as described above using the Euclidean distance for some number of iterations.
    \item For each cluster calculate its cluster centroid (mean) and covariance matrix.
    \item Allocate each point in the dataset to its closest cluster based on the Mahalanobis distance between it and all clusters centroids.
    \item Repeat the above two steps up to a number of times equal to that set at the start.
\end{enumerate}

In all cases, choosing the correct value for the parameter \(K\) is important. For our toy examples in Section \ref{sec:Applications}, we have prior knowledge of the number of clusters. However, we can also make an educated estimate at the number of clusters we should use by running the algorithm for several values of \(K\) and observing the resulting scree plot. 

Choosing a reasonably high value for the number of starts in the algorithm is also key, since the stochastic nature of choosing the initial points to be the cluster centroids means that the quality of the output can often vary greatly \cite[pages 40 - 41]{kassambara2017practical}. 

Since in our situation we know beforehand to what group each data point is supposed to belong, in Section \ref{sec:Applications}, we can also plot graphs and compare metrics to illustrate the performance and accuracy of the clustering from using different distance measures. In the next section though, we investigate the conclusions already drawn in the literature relating to the effectiveness of the Mahalanobis and other distance measures for clustering. 

\section{Critical Assessment}\label{sec:CriticalAssessment}
In this section, we seek to explore the use of different distance measures in cluster analysis. We investigate whether there is any conclusive evidence to support the use of the Mahalanobis distance and other similar distance measures that take correlations into account, over more traditional metrics such as the Euclidean distance. In particular, in Section \ref{subsec:Papers}, we discuss what some papers have found in this area. Then, in Section \ref{subsec:ChatGPT}, we analyse outputs gained from ChatGPT in response to prompts relating to this topic, to assess its knowledge and accuracy in this area.

\subsection{Discussion from Papers}\label{subsec:Papers}
Here, we consider the findings and opinions of a few papers relating to this topic of different distance measures for cluster analysis. The papers we look at either aim to investigate this directly, or we see if we can draw conclusions from their choice for a specific application. We explore whether there is evidence that the Mahalanobis distance is an effective and/or widely used metric for clustering. Additionally, we consider papers that offer alternatives to the Mahalanobis distance, and investigate whether they conclude any strong benefits to these methods.

Firstly, the paper `Simplicial variances, potentials and Mahalanobis distances' from 2018 \cite{PRONZATO2018276} discusses generalisations of the Mahalanobis distance known as `\(k\)-simplicial distances'. These distances are parameterised by \(k \in \{1,\dotsc,d\}\), and the paper states that in the case \(k = d\), it reduces to the usual Mahalanobis distance. The examples section of the paper compares between using the Euclidean, Mahalanobis and \(k\)-simplicial distances for cluster analysis on simulated data. The authors find that their new distance measure introduced does give significant improvements to the outcomes in the case where the intrinsic dimension of the data is lower than the \(d\)-dimensional space to which it belongs. The paper also concludes that the \(k\)-simplicial distance metric seems to benefit from there being more points present in each cluster, in contrast with the other distances considered. Both of these observations suggest that the underlying characteristics of the data to be clustered are extremely important in terms of the effectiveness of a particular distance measure over another. However, the paper does not compare between the different distance measures for an application to a real-life dataset, choosing instead to only use the \(k\)-simplicial distances for this purpose. This means that we cannot conclude from the paper whether this \(k\)-simplicial distance is beneficial in practice.

The paper `Simplicial and Minimal-Variance Distances in Multivariate Data Analysis' from 2022 \cite{gillard2022simplicial} appears in some respects to directly follow on from the above discussed work \cite{PRONZATO2018276}. This is because it further discusses the \(k\)-simplicial distances, providing additional formulae to calculate them in practice. In addition, though, the paper goes further by introducing another class of distance measures known as `\(k\)-minimal-variance distance'. The authors perform detailed explorations into using these two classes of distances alongside the Euclidean and Mahalanobis distances by comparing the outcomes obtained when applying these in the \(K\)-means clustering algorithm to several different datasets. They find that for their application to a small low-dimensional dataset, the Mahalanobis distance gives better results than the Euclidean distance, suggesting the importance of taking correlations into account. However, for a couple of the larger-dimensional datasets used, the authors conclude that the \(k\)-simplicial and \(k\)-minimal-variance distances perform better, critically though, only for appropriate choices of \(k\). It is suggested in the paper that the pseudo-Mahalanobis distance may perform worse due to the intrinsic lower dimensionality of the data, causing singular covariance matrices which then have to be inverted via the Moore-Penrose pseudo-inverse. This implies that considering the rank of the data may be an important factor in choosing the best distance measure. Overall, despite the seemingly positive results found in favour of the recently developed \(k\)-simplicial and \(k\)-minimal-variance distance classes for certain datasets, they seem to be neither further studied nor widely implemented beyond these two papers. Therefore, further research could be done by applying these in more real-life scenarios to deduce if there is a clear benefit for their use.

Next, we consider the paper `K-means with Three different Distance Metrics' from 2013 \cite{singh2013k}. Whilst not considering the Mahalanobis distance specifically, it is of interest to us since it compares the results of using the Euclidean, Manhattan and Minkowski distances with the \(K\)-means clustering algorithm, where we also apply some of these alongside the Mahalanobis distance in Section \ref{sec:Applications}. For their investigations, the authors use simulated data. They find that the Minkowski distance gives similar results to the Euclidean distance, however, the convergence is much slower, suggesting that there is not a clear advantage to choosing this metric. Additionally, the paper finds that using the Manhattan distance produces the worst performance quality and so it concludes that the Euclidean distance is the best of the three metrics considered. However, though the paper emphasises the fact that the choice of distance measure is greatly influential on the clustering quality, it does not appear to mention if there is any effect noticeable on the success of the methods based on the underlying dataset used. Therefore, applying these distance measures to different empirical datasets and investigating if this gives alterations in the results could further extend the conclusions of this work. 

The paper `Wind farm monitoring using Mahalanobis distance and fuzzy clustering' from 2018 \cite{WindFarms} is an example of literature relating to this topic based on a practical application. As part of the paper, the author compares the results generated when using the Euclidean, Manhattan, Minkowski, Maximum and \(\chi^{2}\) distances with those using the Mahalanobis distance. The paper gives reasons as to why the main comparison reduces to between the Euclidean and Mahalanobis distances, and the conclusion is that, overall, the Mahalanobis distance provides the most advantages, being more effective in terms of accuracy than the Euclidean distance for the specific example considered. This suggests that the Mahalanobis distance is indeed used in practice for some applications and suggests that it can have advantages over the Euclidean distribution. On the other hand, we remark that this may be heavily context dependent. For example, the paper `Choices of Distance Matrices in Cluster Analysis: Defining Regions' from 2001 \cite{mimmack2001choice} compares the  use of the Euclidean and Mahalanobis distances. However, for this specific application, it is the Euclidean distance that the authors find to be the better choice, with the Mahalanobis distance not being deemed appropriate for the task. This could suggest that there are conflicting reports in existence as to the benefits and drawbacks of using the Mahalanobis distance. However, we can also note that this paper is quite old now. Therefore, changes such as alterations in computing power could possibly affect future considerations in this area as well. Overall though, it implies that different distance measures should be considered and the results compared to make the best decision for any given scenario, since there appears to be no one optimal solution for all applications.

In summary, we conclude that there appears to be a mix of opinions regarding the efficiency of the Mahalanobis distance in literature, depending on the data and specific scenario being considered. We have also seen that there are alternatives to and generalisations of the Mahalanobis distance that may appear to work well in certain situations. However, these do not yet seem to be in widespread use, possibly due to their recency of development or their increased computational complexity. Of the Euclidean, Manhattan, Minkowski and Maximum distances, it would seem that the Euclidean distance is frequently the preferred method. On the other hand, there seems to be no clear favourite between the Euclidean and Mahalanobis distance for clustering algorithms, with both seeming to be used in practice. Thus, it appears to be important to consider many factors of the underlying dataset such as the size, intrinsic dimensionality, and the importance of correlations when deciding on a distance measure. In particular, it seems to be beneficial to experiment with different distance measures, and compare the results seen for each, before making a choice of distance metric for a particular application.

\subsection{ChatGPT}\label{subsec:ChatGPT}
We now assess information obtained from ChatGPT in response to prompts related to distance measures for cluster analysis, with a particular emphasis on the Mahalanobis distance, and its comparison with the Euclidean distance. This will enable us to gain an understanding of its awareness of this topic, and also allow us to evaluate its accuracy by comparing the answers received with what we have observed from literature. 

The first question asked to ChatGPT was to gauge an appreciation of whether it felt there was a particular distance measure that is always best for use with cluster analysis on numerical data. The specific prompt used was
\begin{BoxB}
    {\small``What is the best distance measure for cluster analysis on numerical data?''}
\end{BoxB}
which resulted in the answer given in the box below \cite{ChatGPT}. The response gained here is encouraging since it highlights some of the many different distance measures possible, some of which we have already discussed, and it does not try and claim that there is a single best distance metric. Comparing the definitions given here to those we presented in Section \ref{subsec:DistanceMeasures}, it would seem that the information it presents is fairly accurate in this case, if not completely precise. This may be since cluster analysis is a reasonably classical technique in multivariate data analysis, thus there has been much research done in this area that could contribute to its knowledge base. The statement that in deciding which distance measure to use the characteristics of the data should be taken into account also agrees with the conclusions we found in Section \ref{subsec:Papers}.
\begin{BoxB}
    {\small``The choice of distance measure for cluster analysis on numerical data depends on various factors, including the nature of the data, the distribution of the data, and the characteristics of the clusters you expect to find. Here are some commonly used distance measures for cluster analysis on numerical data:

1. \textbf{Euclidean Distance}: This is the most commonly used distance measure and is suitable for data with continuous variables. It calculates the straight-line distance between two points in Euclidean space.

2. \textbf{Manhattan Distance (City Block Distance)}: This distance is calculated as the sum of the absolute differences between the coordinates of two points. It is useful when dealing with data that are not normally distributed or when outliers are present.

3. \textbf{Chebyshev Distance}: Also known as maximum metric distance, it calculates the maximum absolute difference between the coordinates of two points along any dimension. It is robust to outliers and can handle data with varying scales.

4. \textbf{Minkowski Distance}: This is a generalization of both the Euclidean and Manhattan distances. By adjusting the parameter \( p \), you can make it behave like either of them. When \( p = 2 \), it is equivalent to the Euclidean distance, and when \( p = 1 \), it is equivalent to the Manhattan distance.

5. \textbf{Mahalanobis Distance}: This distance measure takes into account the variability of the data by considering the covariance matrix. It is useful when dealing with data with correlated features and different variances.

6. \textbf{Cosine Similarity}: Instead of measuring the distance between points, cosine similarity measures the cosine of the angle between two vectors. It is often used for text data or high-dimensional data where the magnitude of the vectors is not relevant.

7. \textbf{Correlation Distance}: This measures the dissimilarity between two vectors as \(1 - \text{correlation coefficient}\). It is useful when the direction of variation matters more than the magnitude, such as in gene expression data or customer preferences.

The choice of distance measure should be made based on the characteristics of your data and the goals of your analysis. It's often a good practice to try multiple distance measures and clustering algorithms to see which combination produces the most meaningful results for your specific dataset and objectives.''}
\end{BoxB}

The next idea we explored was to try to understand how often it would appear that the Mahalanobis and Euclidean distances are used in practice. From Section \ref{subsec:Papers}, we might expect the more basic Euclidean distance to be used more routinely. However, we have also seen evidence that in certain scenarios the Mahalanobis distance can be more effective, and it is indeed implemented for some applications. We use the separate prompts 
\begin{BoxB}
    {\small``How frequently is the Euclidean distance used for cluster analysis in practice?''}
\end{BoxB}
and 
\begin{BoxB}
    {\small``How frequently is the Mahalanobis distance used for cluster analysis in practice?''}
\end{BoxB}
to obtain, respectively, the answers given in the boxes below \cite{ChatGPT}.
\begin{BoxB}
    {\small``The Euclidean distance is one of the most commonly used distance metrics in cluster analysis. It measures the straight-line distance between two points in a Euclidean space. Cluster analysis, which involves grouping data points into clusters based on similarity, often relies on distance metrics to determine the similarity or dissimilarity between data points.

In practice, the Euclidean distance is frequently used because of its simplicity, intuitive interpretation, and applicability to a wide range of data types, including numerical and continuous variables. It is particularly popular in methods such as k-means clustering, hierarchical clustering, and other distance-based clustering algorithms.

However, it's essential to note that the choice of distance metric depends on the nature of the data and the specific requirements of the analysis. In some cases, alternative distance metrics such as Manhattan distance, Mahalanobis distance, or cosine similarity may be more appropriate, especially for non-Euclidean data spaces or when dealing with high-dimensional data.

Overall, while the Euclidean distance is widely used in practice for cluster analysis, researchers and practitioners should carefully consider the characteristics of their data and the objectives of their analysis when selecting a distance metric.''}
\end{BoxB}
\begin{BoxB}
    {\small``The Mahalanobis distance is a widely used metric in cluster analysis, especially in situations where the clusters are not necessarily spherical or where the data may be correlated. However, its frequency of use can vary depending on the specific context, the nature of the data, and the objectives of the analysis.

In practice, the Mahalanobis distance is often employed in various fields such as:

1. \textbf{Data mining and machine learning}: It is commonly used in clustering algorithms such as k-means clustering, Gaussian mixture models, and hierarchical clustering. In these algorithms, the Mahalanobis distance helps to measure the dissimilarity between data points and cluster centroids while taking into account the covariance structure of the data.

2. \textbf{Pattern recognition}: Mahalanobis distance is utilized for classification tasks, anomaly detection, and feature selection, especially when dealing with high-dimensional data where feature correlations need to be considered.

3. \textbf{Quality control and process monitoring}: It is employed to assess the similarity of observations in multivariate data sets, enabling the detection of outliers or deviations from expected patterns.

4. \textbf{Biostatistics and bioinformatics}: Mahalanobis distance is used in analyzing biological data, such as gene expression data, protein-protein interaction networks, and epidemiological studies, to identify clusters or patterns of interest.

5. \textbf{Image processing and computer vision}: It finds applications in image segmentation, object recognition, and tracking, where the distance metric needs to account for the correlation between image features.

While the Mahalanobis distance offers advantages over other distance metrics in certain scenarios, such as handling correlated variables and providing a more meaningful measure of dissimilarity, its usage may be less frequent in simpler clustering tasks or when the assumption of multivariate normality is not met. In such cases, alternative distance metrics like Euclidean distance or Manhattan distance may be preferred. Ultimately, the choice of distance metric depends on the characteristics of the data and the specific goals of the analysis.''}
\end{BoxB}
These responses, whilst not directly answering the questions, seem to agree with what we would expect. For example, it claims that the Euclidean distance is one of the most popular distance measures, though the Mahalanobis distance can be beneficial particularly in certain industries. A key part of the responses though, are that they reiterate the fact that not all distance measures are suited to all applications. This highlights the importance of careful research into the data for a specific situation before selecting a distance measure, possibly even using several first and comparing their qualities before making a final decision. We also remark though that, in both responses it recommends the Manhattan distance as a possible alternative. However, one of the papers we discussed in Section \ref{subsec:Papers} found this to be less effective than the Euclidean distance. Therefore, further research into this for different datasets could be performed to improve our understanding here. 

The final exploration we perform here is to investigate what some of the benefits of the Euclidean and Mahalanobis distances are, as well as in what situations they may be less useful. We again split this into the two prompts
\begin{BoxB}
    {\small``What are the advantages of using the Euclidean distance for cluster analysis and when might it fail?''}
\end{BoxB}
and
\begin{BoxB}
    {\small``What are the advantages of using the Mahalanobis distance for cluster analysis and when might it fail?''}
\end{BoxB}
The responses obtained can be seen respectively in the boxes below \cite{ChatGPT2}.
\begin{BoxB}
    {\small``Euclidean distance is commonly used in cluster analysis due to its simplicity and ease of computation. Some advantages of using Euclidean distance for cluster analysis include:

1. \textbf{Ease of Computation}: Euclidean distance is straightforward to calculate and requires only basic arithmetic operations. It measures the straight-line distance between two points in Euclidean space, which is intuitive and easy to understand.

2. \textbf{Geometric Interpretation}: Euclidean distance corresponds to the length of the straight line segment that connects two points in space. This geometric interpretation is useful for visualizing clusters and understanding their spatial relationships.

3. \textbf{Widely Applicable}: Euclidean distance can be used in various types of data, including numerical, continuous, and categorical variables (if properly encoded). This versatility makes it suitable for a wide range of clustering tasks.

4. \textbf{Implementation Availability}: Many clustering algorithms, such as K-means, hierarchical clustering, and DBSCAN, utilize Euclidean distance as a similarity measure. These algorithms are widely available in software libraries and frameworks, making it convenient for practitioners to apply Euclidean distance in cluster analysis.

However, Euclidean distance may fail or exhibit limitations in certain scenarios:

1. \textbf{Sensitive to Scale}: Euclidean distance is sensitive to the scale of features. If the features have different units or scales, it can lead to biased results. Preprocessing techniques like standardization or normalization can mitigate this issue to some extent.

2. \textbf{Curse of Dimensionality}: In high-dimensional spaces, Euclidean distance tends to lose its discriminative power, as distances between points become increasingly similar. This phenomenon, known as the curse of dimensionality, can affect the quality of clustering results, particularly when dealing with high-dimensional data.

3. \textbf{Assumption of Linear Separability}: Euclidean distance implicitly assumes linear separability, meaning that clusters are well-separated by straight-line distances. In reality, clusters may exhibit complex shapes or non-linear relationships, where Euclidean distance may not accurately capture the underlying structure.

4. \textbf{Outliers Sensitivity}: Euclidean distance is sensitive to outliers, as it considers all dimensions equally. Outliers can disproportionately influence the distance metric and potentially distort the clustering results.

In summary, while Euclidean distance offers simplicity and ease of use in cluster analysis, practitioners should be mindful of its limitations, particularly regarding scale sensitivity, high-dimensional data, non-linear relationships, and outliers. It's essential to evaluate the appropriateness of Euclidean distance based on the characteristics of the data and the objectives of the clustering task.''}
\end{BoxB}
\begin{BoxB}
    {\small``The Mahalanobis distance is a metric used to measure the distance between a point and a distribution. It takes into account the covariance structure of the data, making it particularly useful in cases where the data may be correlated or have different variances along different dimensions. Here are some advantages of using the Mahalanobis distance for cluster analysis:

1. \textbf{Accounts for covariance}: Unlike Euclidean distance, which treats each dimension independently, Mahalanobis distance considers the covariance between dimensions. This makes it effective for datasets where variables are correlated.

2. \textbf{Sensitive to scale}: Mahalanobis distance is sensitive to the scale of the variables. This means it can handle datasets with variables measured in different units without requiring normalization.

3. \textbf{Optimal for normally distributed data}: When the data is normally distributed, Mahalanobis distance provides an optimal measure of dissimilarity between points.

4. \textbf{Useful for elliptically shaped clusters}: It is effective in identifying clusters that are not necessarily spherical but rather have elliptical shapes.

However, there are situations where using Mahalanobis distance might fail or be less effective:

1. \textbf{Non-Gaussian distributions}: Mahalanobis distance assumes multivariate normality. If the data distribution significantly deviates from normality, Mahalanobis distance may not provide accurate results.

2. \textbf{Small sample sizes}: In cases where the sample size is small relative to the number of variables, estimating the covariance matrix accurately becomes challenging, leading to unreliable Mahalanobis distances.

3. \textbf{Outliers}: Mahalanobis distance can be sensitive to outliers, especially if they are not accounted for properly. Outliers can disproportionately affect the estimation of the covariance matrix, leading to biased distance measurements.

4. \textbf{High dimensionality}: In high-dimensional spaces, estimating the covariance matrix accurately becomes increasingly difficult, which can result in unreliable Mahalanobis distances.

In summary, while Mahalanobis distance offers several advantages for cluster analysis, such as accounting for covariance and variable scales, its effectiveness can be compromised in situations where data distributions deviate from normality, sample sizes are small, outliers are present, or in high-dimensional spaces. Therefore, it's essential to consider these factors when applying Mahalanobis distance for cluster analysis.''}
\end{BoxB}
The responses we have obtained here are quite similar to the ones we have received previously in some respects. This may suggest that its knowledge in this area is limited, being only able to repeatedly give basic summary information and not able to give specific details. However, this may be an advantage, since it encourages the user to investigate the characteristics of their dataset before making a decision. Some of the advantages it discusses for both distance measures align with what we have seen previously. For example, the Euclidean distance is widely understood and simple to compute, whilst one of the key benefits of the Mahalanobis distance mentioned in the papers we considered is the fact that it takes the covariance matrix into account. On the other hand, we may question its reliability and/or the clarity of its expression at times. This is because it states that both the Euclidean and Mahalanobis distances are sensitive to the scale of the data. However, it claims that this is an advantage in one case but a disadvantage in the other. A common theme in all the responses obtained seems to be that ChatGPT gives much information of the subject matter of the prompt, but it does not answer the question directly, instead encouraging the user to consider some of the points raised to make their own decision. This could also suggest that there are very few definitive answers when it comes to distance measures for clustering algorithms so that it is unable to offer advice in this area.

Overall, we may conclude that ChatGPT seems to have a basic understanding of the different distance measures possible for use in cluster analysis. In particular, it emphasises the important fact that there appears to be no clear `best' metric, with all decisions needing to be made specific to the data. This agrees with what we have observed in Section \ref{subsec:Papers}. However, we can also remark that some of its responses seem contradictory at times. This means that it should be used with caution and it highlights the importance of checking the information gained from it against other more reliable sources. In the next section we move on to applying some of the discussed distance measures to simulated and real-life datasets.

\section{Applications}\label{sec:Applications}
In this section, we present some investigations using the different distance measures defined in Section \ref{subsec:DistanceMeasures} with the K-means algorithm outlined in Section \ref{subsec:KMeansClustering} for cluster analysis. We start in Section \ref{subsec:SimulatedExample} by simulating some two-dimensional data to use. Then, in Section \ref{subsec:DatasetExample}, we apply our clustering methods to subsets of the \emph{Dry Bean} dataset \cite{misc_dry_bean_dataset_602} available from the UCI Machine Learning Repository. For this example, we reduce the dimension of the dataset to two using Principal Component Analysis in order to plot the results. Note that in both cases we already know the desired clustering. Hence, we can compare the performance of the different distance measures by observing how accurate the resulting clustering is. The code for all this is completed using R and it can be seen in Appendix \ref{sec:Appendix}.

\subsection{Simulated Dataset Example}\label{subsec:SimulatedExample}
\begin{figure}[ht]
    \centering
    \begin{minipage}{.45\textwidth}
        \centering
    \includegraphics[width=\linewidth]{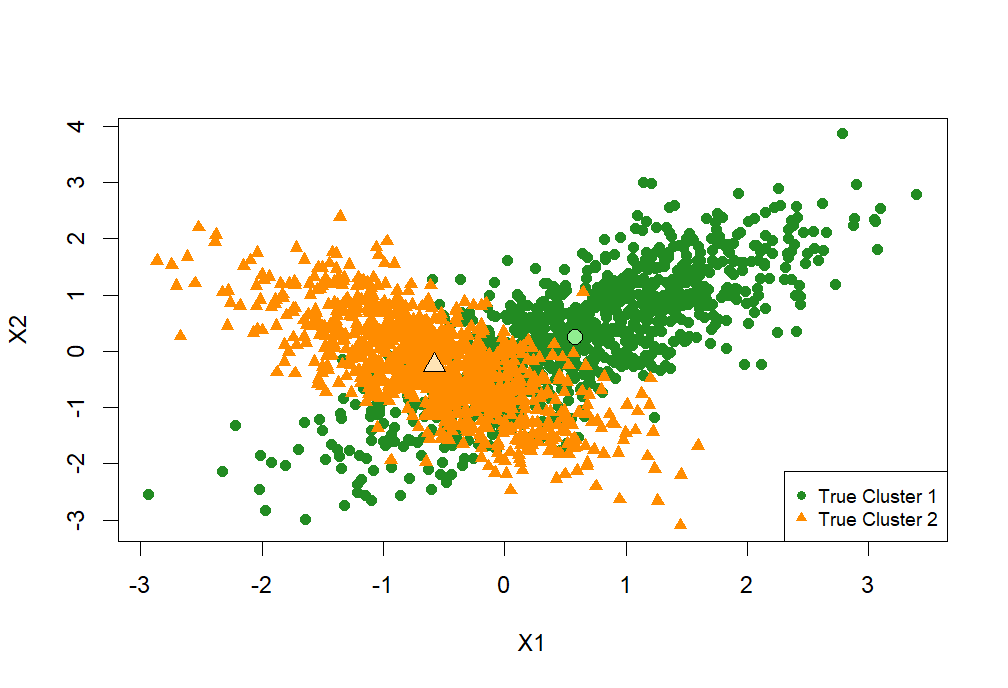}
    \caption{\textsl{Scatter plot showing the two clusters and cluster centres \\for the scaled simulated data.}}
    \label{fig:TrueScatter}
    \end{minipage}
    \begin{minipage}{.45\textwidth}
        \centering
    \includegraphics[width=\linewidth]{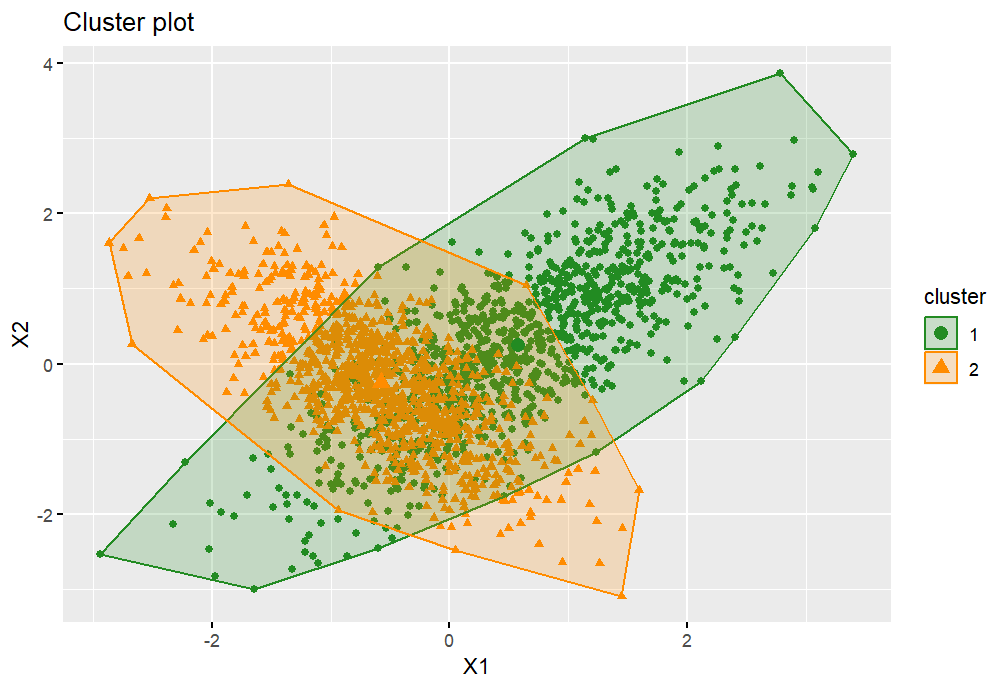}
    \caption{\textsl{Cluster plot when using the true clusters \\for the scaled simulated data.}}
    \label{fig:TrueCluster}
    \end{minipage}
\end{figure}

Here we present  a toy example using simulated data to explore the effect of using different distance measures in the K-means algorithm for clustering. To produce the data, we generate 1000 points from each of two different multivariate normal distributions. Specifically, we take 1000 points from the distribution
\begin{equation*}
    \mathcal{N}\left(\left(\begin{array}{c}
        1\\
        1
    \end{array}\right),\left(\begin{array}{cc}
        1.5 &  1\\
        1 & 1
    \end{array}\right)\right),
\end{equation*}
and 1000 more from the distribution
\begin{equation*}
    \mathcal{N}\left(\left(\begin{array}{c}
        -0.5\\
        0.5
    \end{array}\right),\left(\begin{array}{cc}
        0.8 &  -0.5\\
        -0.5  & 0.6
    \end{array}\right)\right).
\end{equation*}
We label the former set of points with a 1 and the latter with a 2, then join the sets together to get a dataset of 2000 points in \(\mathbb{R}^{2}\). These labels then form the `true' clusters we aim to reproduce using the overall dataset. Since the clustering algorithm can be sensitive to the scale of the data, the next step is to standardise the combined dataset so that the two variables have mean zero and variance one. At this stage, we can calculate what the true sample cluster centroids (means) are for our scaled sample data. We find that they are approximately (0.579, 0.248) and (-0.579, -0.248) respectively, to three decimal places. A scatter plot showing the scaled data and cluster centroids can be seen in Figure \ref{fig:TrueScatter}. Additionally, Figure \ref{fig:TrueCluster} illustrates the appearance of the cluster plot given the true clustering scheme. From these two figures, we can see that there is a moderate amount of overlap in the centre where there are points belonging to both clusters, though distinct clusters do appear to be visible. Additionally, we can see that the clusters are highly elliptical, suggesting that the variables have a significant amount of correlation within the clusters. From our understanding using the previous sections, we may expect that this will result in the Mahalanobis distance possibly working well in this scenario. However, this is something we now progress to investigate. 

\begin{figure}[ht]
    \centering
    \includegraphics[width=0.45\linewidth]{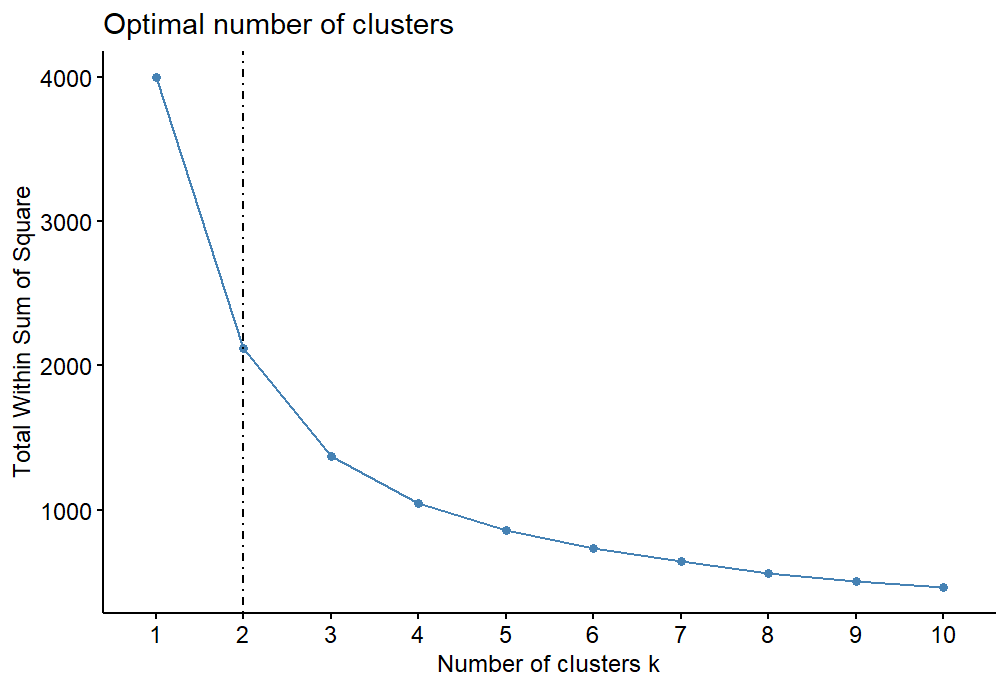}
    \caption{\textsl{Scree plot for the scaled simulated dataset.}}
    \label{fig:ScreePlot}
\end{figure}

In this situation, we know that the true number of clusters is two, therefore we choose \(K = 2\) in the \(K\)- means algorithm. However, for completeness we also choose to illustrate the scree plot that could be used in practice to determine the optimal number of clusters. This is shown in Figure \ref{fig:ScreePlot}. The \(y\)-axis displays the total within cluster sum of squares for the Euclidean distance. From the figure, we can see that there is a large jump in the total within cluster sum of squares value at \(K = 2\), confirming our choice. However, we also remark that given only this figure, it may seem as if three clusters could also be a good choice. Thus, for any given application, it may be a good idea to try different values of \(K\) and compare the performances in order to decide the optimal number of clusters.

\subsubsection{Euclidean Distance}
\begin{figure}[ht]
    \centering
    \begin{minipage}{.45\textwidth}
        \centering
        \includegraphics[width=1\linewidth]{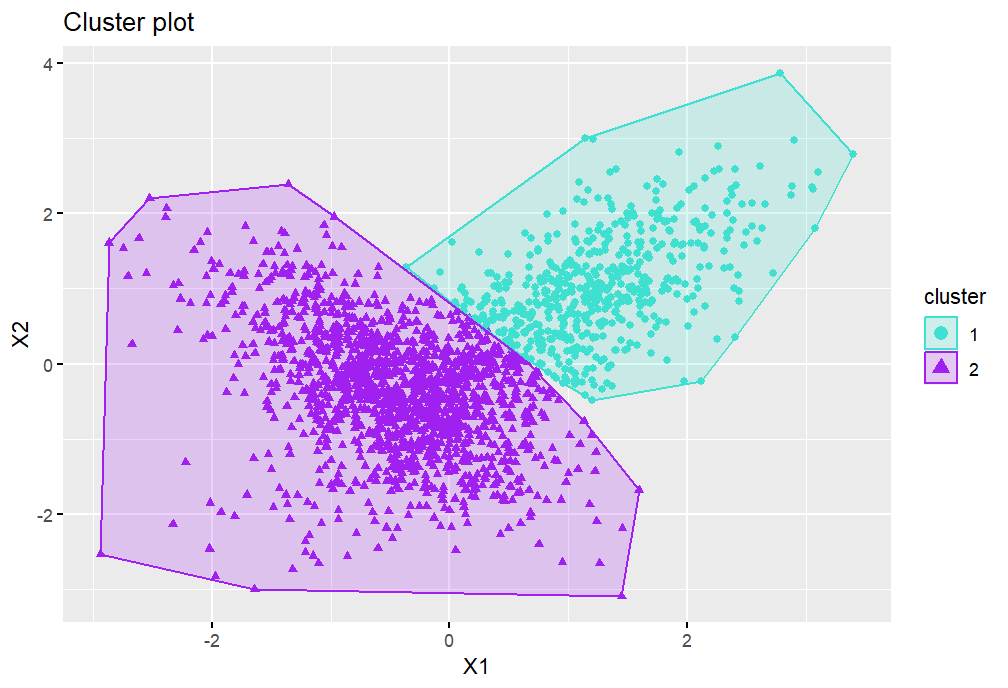}
        \caption{\textsl{Cluster plot showing the clusters produced following the application of the \(K\)-means algorithm with the Euclidean distance on the scaled simulated dataset.}}
        \label{fig:EuclideanCluster}
    \end{minipage}
    \begin{minipage}{.45\textwidth}
        \centering
        \includegraphics[width=1\linewidth]{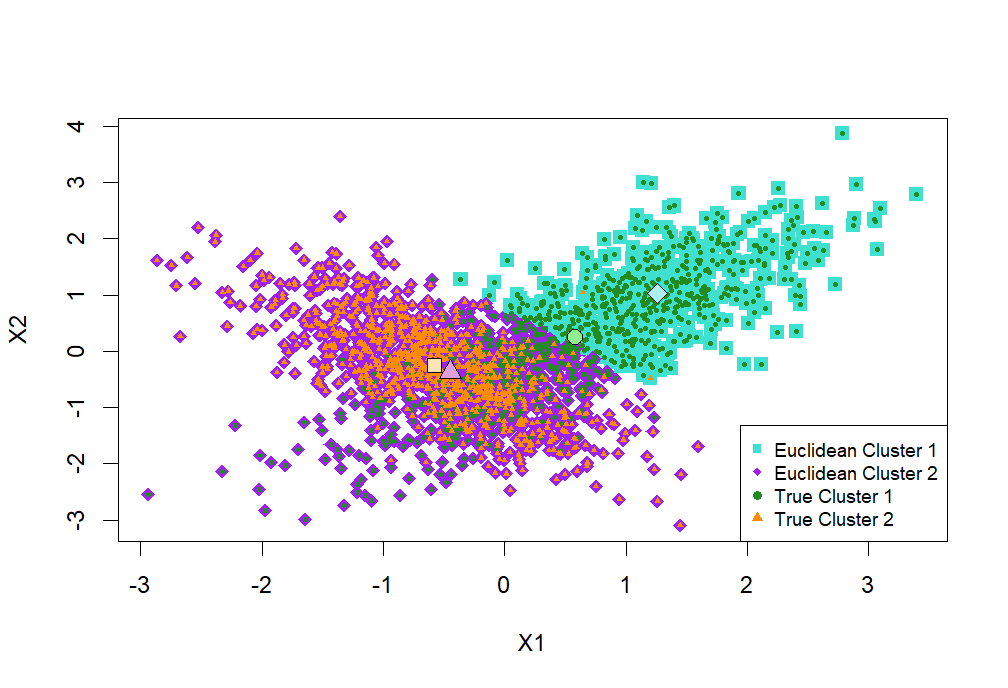}
        \caption{\textsl{Scatter plot showing the true clusters on top of the clusters generated using the Euclidean distance in the \(K\)-means algorithm on the scaled simulated dataset.}}
        \label{fig:EuclideanScatter}
    \end{minipage}
\end{figure}

\begin{figure}[ht]
    \centering
        \includegraphics[width=0.45\linewidth]{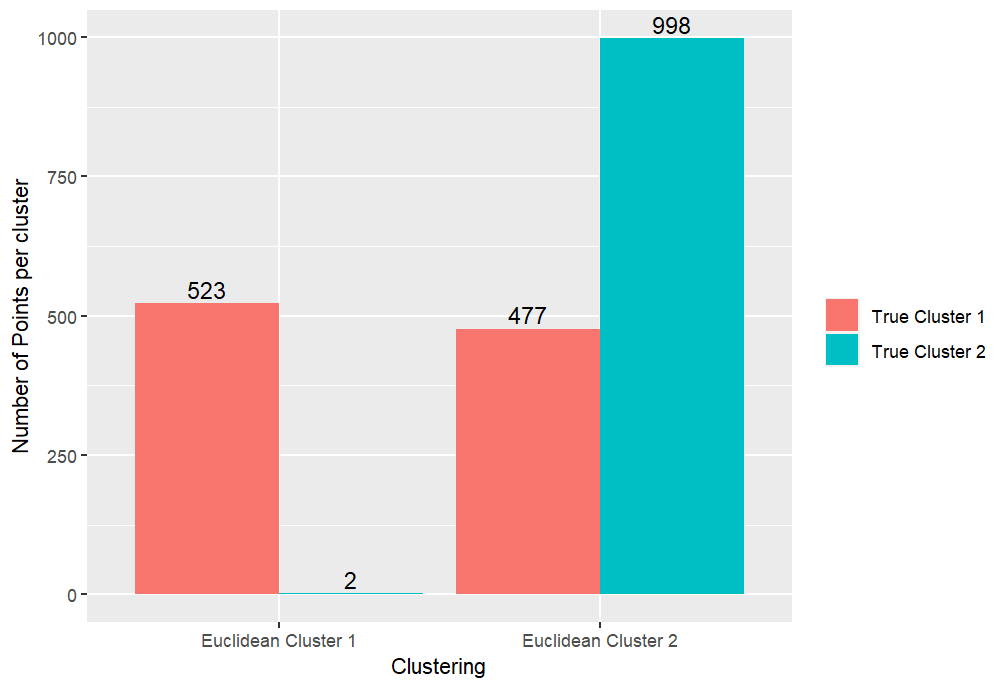}
        \caption{\textsl{Bar plot showing the number of misclassified and correctly classified points in each cluster when using the Euclidean distance in the \(K\)-means algorithm on the scaled simulated dataset.}}
        \label{fig:EuclideanBar}
\end{figure}

We now progress to executing the \(K\)-means algorithm with the Euclidean, Manhattan, Maximum and Mahalanobis distances on the scaled simulated dataset. Firstly, we use the traditional Euclidean distance. A graphical representation of the clustering generated can be seen in Figure \ref{fig:EuclideanCluster}. In order to compare the accuracy of the clustering with the true clustering, we also create the scatter plot that can be seen in Figure \ref{fig:EuclideanScatter}. Here, the points assigned to cluster 1 using the Euclidean distance are coloured turquoise whilst those assigned to cluster 2 are given by the purple diamonds. On top of this we plot the true clustering, so that orange on purple indicates a correct classification, whereas green on purple highlights a misclassification of a group 1 point into cluster 2. Similarly, green on turquoise is the correct assignment of cluster 1 points, whereas orange on green indicates a true group 2 point being clustered into group 1. In addition, we plot the true and Euclidean-distance-generated cluster centroids, indicated by the larger, paler, black-outlined points of the corresponding shape for each cluster. Observing the plot, we can see that the centroids for the true and generated cluster 2 are reasonably similar. However, it would appear that the Euclidean cluster 2 incorporates much of the true cluster 1, since the Euclidean centroid for cluster 1 is some distance from the true centroid. In order to assess the accuracy of the clusters numerically, we can use Table \ref{tab:Euclidean}, which summarises the number of points assigned correctly or incorrectly to each cluster. Figure \ref{fig:EuclideanBar} shows a clustered bar plot using this information to highlight the number of points misclassified by the clustering. From these we can see that the Euclidean-distance-based clustering has assigned almost all points in group 2 correctly. However, almost half of the points actually in group 1 have been incorrectly classified as cluster 2, which is less than desirable. We can now use this as a benchmark to compare to the other distance measures.

\begin{table}[ht]
    \centering
    \begin{tabular}{|c|c|c|}
        \hline
         & Euclidean Cluster 1 & Euclidean Cluster 2 \\
        \hline 
        True Cluster 1 & 523 & 477 \\
        \hline
        True Cluster 2 & 2 & 998\\
        \hline
    \end{tabular}
    \caption{\textsl{Tabular representation of the correctly and incorrectly classified points following the application of the \(K\)-means algorithm with the Euclidean distance on the scaled simulated dataset.}}
    \label{tab:Euclidean}
\end{table}

\subsubsection{Manhattan Distance}
\begin{figure}[ht]
    \centering
    \begin{minipage}{.45\textwidth}
        \centering
        \includegraphics[width=1\linewidth]{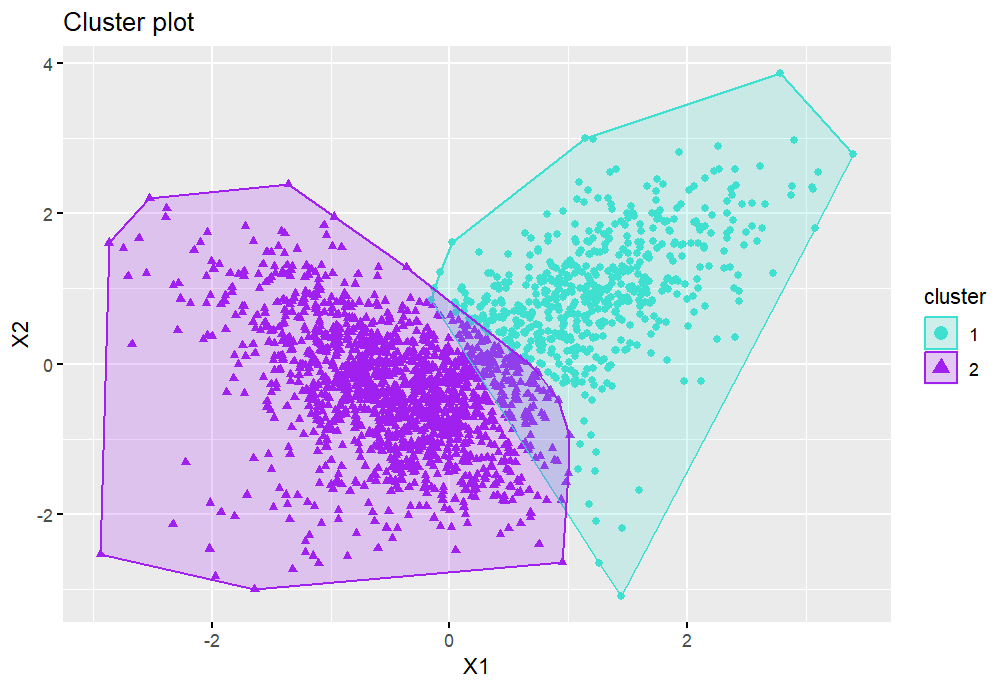}
        \caption{\textsl{Cluster plot showing the clusters produced following the application of the \(K\)-means algorithm with the Manhattan distance on the scaled simulated dataset.}}
        \label{fig:ManhattanCluster}
    \end{minipage}
    \begin{minipage}{.45\textwidth}
        \centering
        \includegraphics[width=1\linewidth]{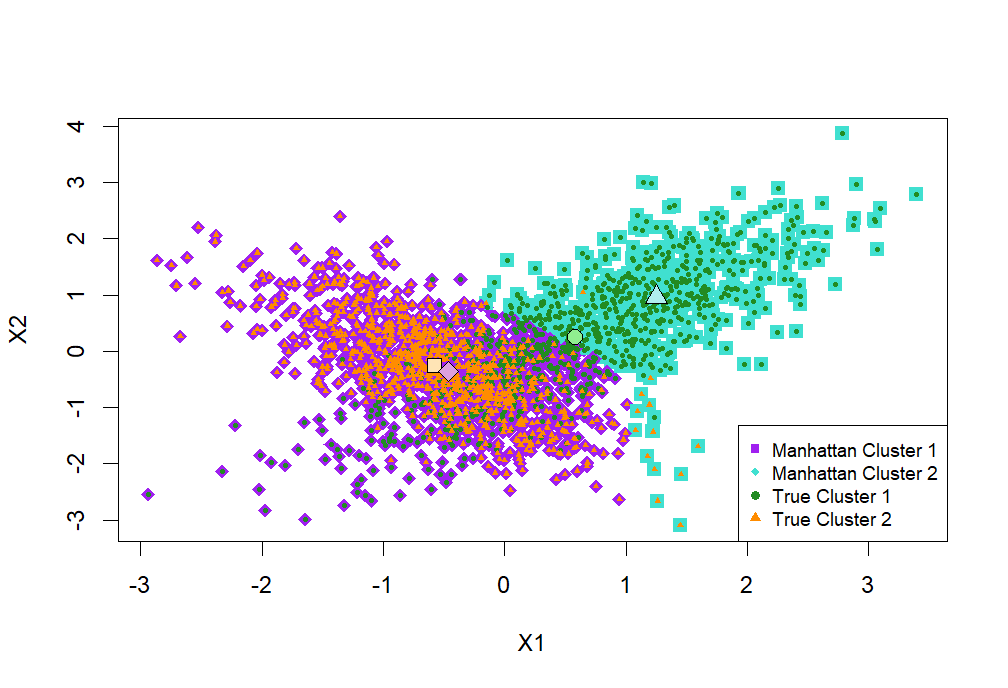}
        \caption{\textsl{Scatter plot showing the true clusters on top of the clusters generated using the Manhattan distance in the \(K\)-means algorithm on the scaled simulated dataset.}}
        \label{fig:ManhattanScatter}
    \end{minipage}
\end{figure}

\begin{figure}[ht]
    \centering
        \includegraphics[width=0.45\linewidth]{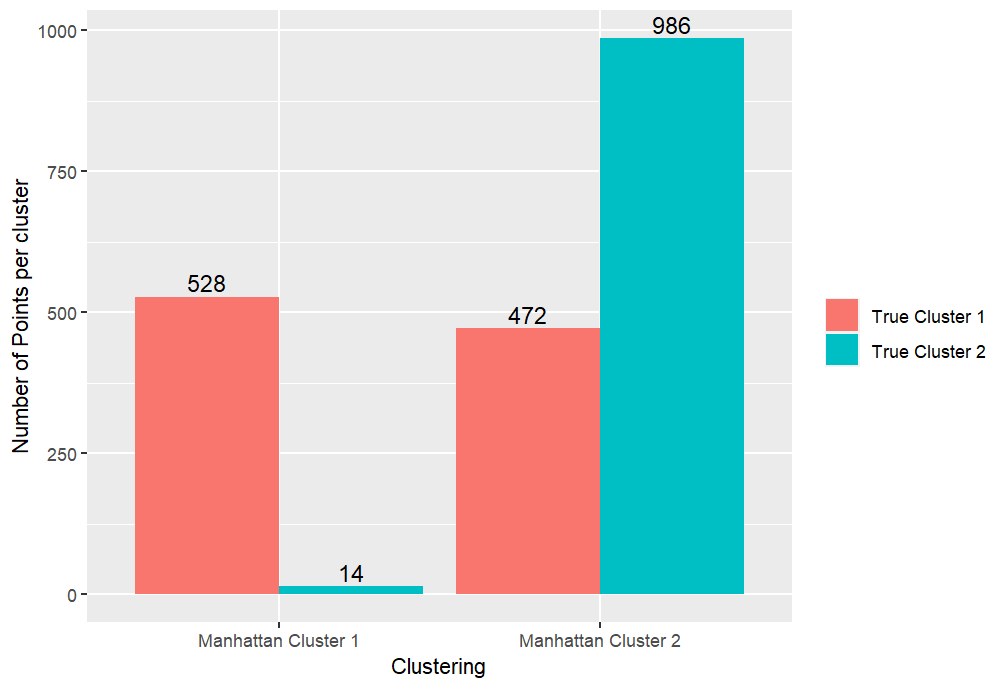}
        \caption{\textsl{Bar plot showing the number of misclassified and correctly classified points in each cluster when using the Manhattan distance in the \(K\)-means algorithm on the scaled simulated dataset.}}
        \label{fig:ManhattanBar}
\end{figure}

Here, we now repeat all of the above experiments but this time use the Manhattan distance instead. We obtain the cluster and scatter plots shown in Figures \ref{fig:ManhattanCluster} and \ref{fig:ManhattanScatter} respectively. From these we can see that the clusters found appear to be very similar to the Euclidean case, with just a few points in the bottom right seeming to switch from being assigned to cluster 2 to cluster 1. Furthermore, we can observe that the cluster centroids appear to be placed almost identically. Considering the summary of misclassified results shown in both Table \ref{tab:Manhattan} and Figure \ref{fig:ManhattanBar}, we can see that the performance of the Manhattan distance is slightly worse than the Euclidean case, with seven more points being incorrectly classified. This time, however, a few more points from the true cluster 2 have been misclassified and fewer points in group 1 have been incorrectly classified compared to the Euclidean case. This slight loss in quality suggests that for this specific case, the more easily interpreted Euclidean distance should be preferred over the Manhattan distance.

\begin{table}[ht]
    \centering
    \begin{tabular}{|c|c|c|}
        \hline
         & Manhattan Cluster 1 & Manhattan Cluster 2 \\
        \hline 
        True Cluster 1 & 528 & 472 \\
        \hline
        True Cluster 2 & 14 & 986\\
        \hline
    \end{tabular}
    \caption{\textsl{Tabular representation of the correctly and incorrectly classified points following the application of the \(K\)-means algorithm with the Manhattan distance on the scaled simulated dataset.}}
    \label{tab:Manhattan}
\end{table}

\subsubsection{Maximum Distance}
\begin{figure}[ht]
    \centering
    \begin{minipage}{.45\textwidth}
        \centering
        \includegraphics[width=1\linewidth]{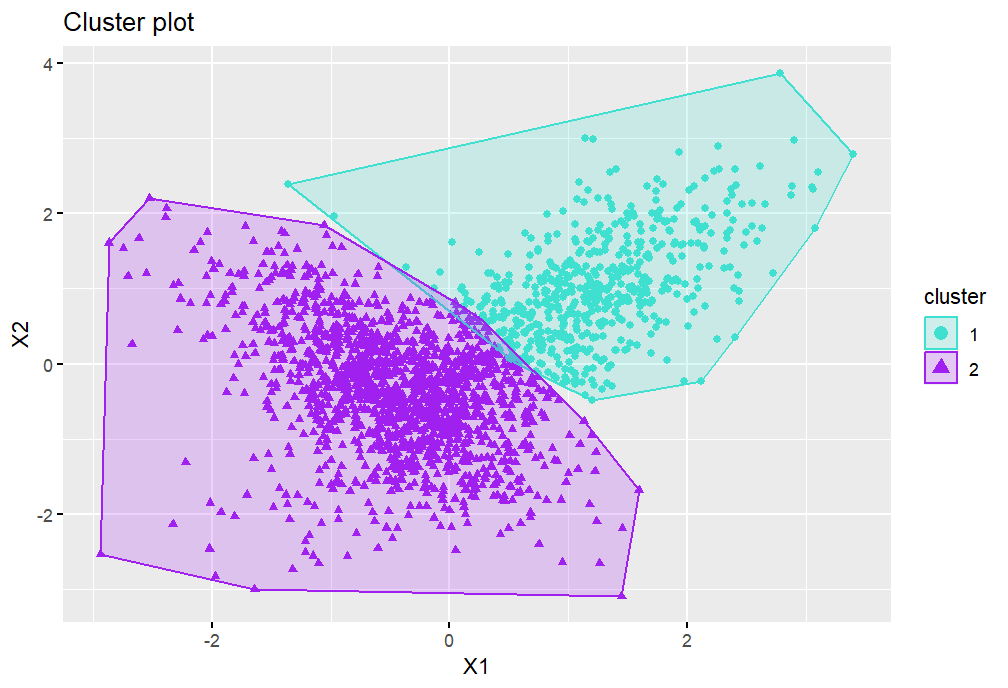}
        \caption{\textsl{Cluster plot showing the clusters produced following the application of the \(K\)-means algorithm with the Maximum distance on the scaled simulated dataset.}}
        \label{fig:MaximumCluster}
    \end{minipage}
    \begin{minipage}{.45\textwidth}
        \centering
        \includegraphics[width=1\linewidth]{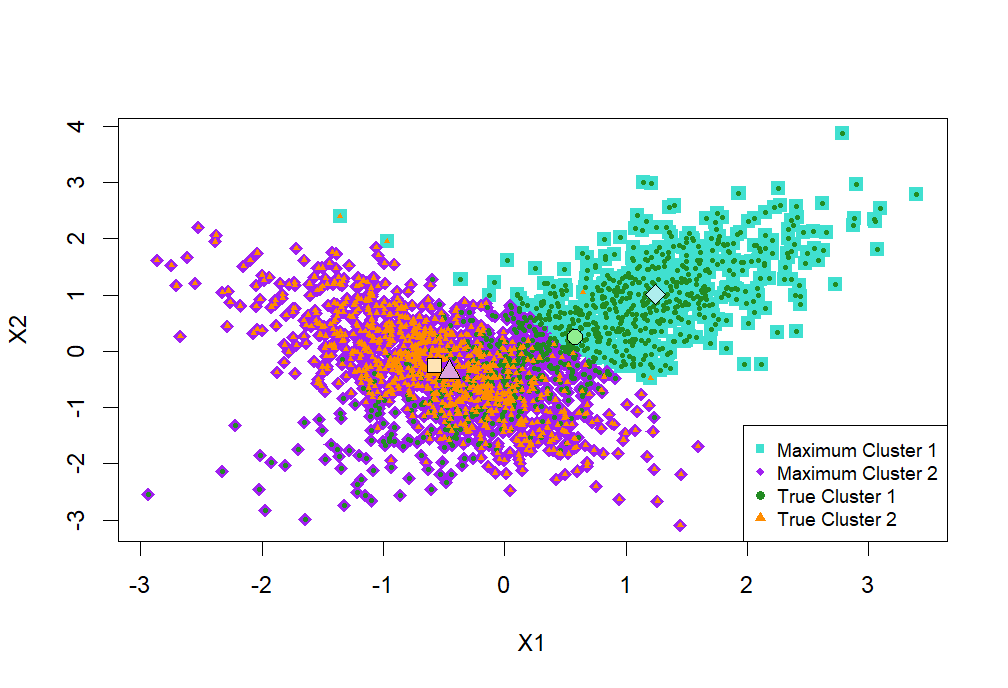}
        \caption{\textsl{Scatter plot showing the true clusters on top of the clusters generated using the Maximum distance in the \(K\)-means algorithm on the scaled simulated dataset.}}
        \label{fig:MaximumScatter}
    \end{minipage}
\end{figure}

\begin{figure}[ht]
    \centering
        \includegraphics[width=0.45\linewidth]{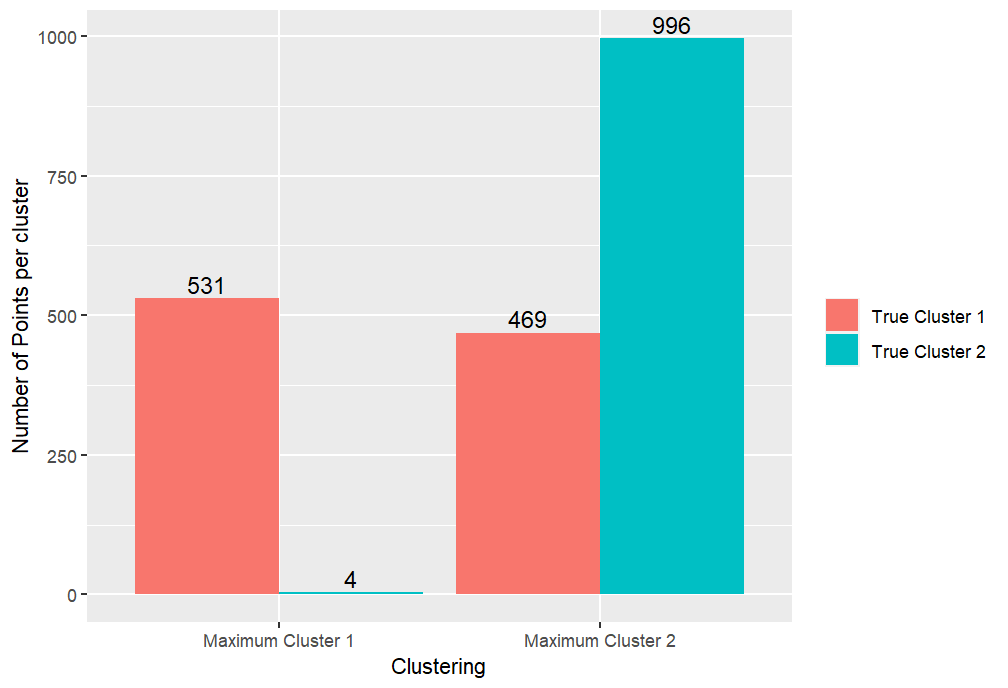}
        \caption{\textsl{Bar plot showing the number of misclassified and correctly classified points in each cluster when using the Maximum distance in the \(K\)-means algorithm on the scaled simulated dataset.}}
        \label{fig:MaximumBar}
\end{figure}

Next, we progress to considering the clustering produced when using the Maximum distance measure. Again, the cluster and scatter plots can be seen in Figures \ref{fig:MaximumCluster} and \ref{fig:MaximumScatter} respectively. These seem much more comparable to the results when using the Euclidean distance when compared to those for the Manhattan distance. However, as has been the case for the other two distance measures, it has failed to encapsulate the elongated intersecting ellipse shapes of the true clusters. Table \ref{tab:Maximum} and Figure \ref{fig:MaximumBar} show the misclassification results from using the Maximum distance. Comparing to the corresponding table and Figure for the Euclidean distance, we can see that there are actually six fewer points misclassified in this case, since although a couple more of group 2 have been incorrectly assigned, more of group 1 have been correctly classified in this case. However, reconsidering Figures \ref{fig:EuclideanCluster} and \ref{fig:MaximumCluster} we might suggest that the former of the Euclidean distance has got the better approximation of the true shape than the latter, due to the turquoise far outlying points in the top left of Figure \ref{fig:MaximumCluster}. We also note that the values for the correctly and incorrectly assigned points are very close to each other, thus in this instance, there would seem no strong evidence to suggest that using the Maximum distance measure would give marked improvements over the more classical and widely understood Euclidean distance.

\begin{table}[ht]
    \centering
    \begin{tabular}{|c|c|c|}
        \hline
         & Maximum Cluster 1 & Maximum Cluster 2 \\
        \hline 
        True Cluster 1 & 531 & 469 \\
        \hline
        True Cluster 2 & 4 & 996\\
        \hline
    \end{tabular}
    \caption{\textsl{Tabular representation of the correctly and incorrectly classified points following the application of the \(K\)-means algorithm with the Maximum distance on the scaled simulated dataset.}}
    \label{tab:Maximum}
\end{table}

\subsubsection{Mahalanobis Distance}
\begin{figure}[ht]
    \centering
    \begin{minipage}{.45\textwidth}
        \centering
        \includegraphics[width=1\linewidth]{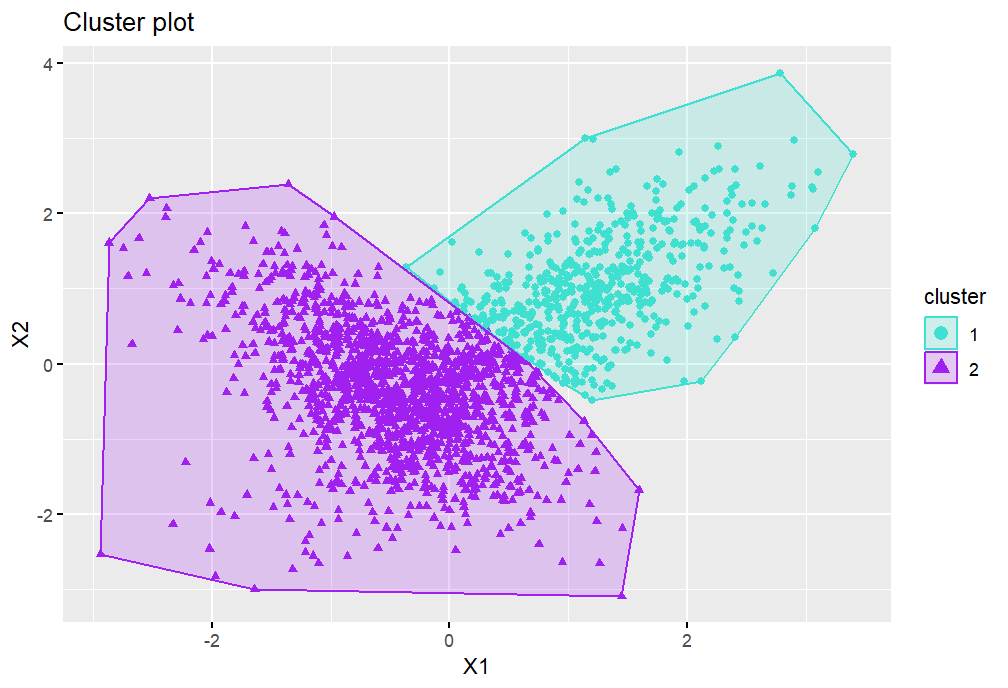}
        \caption{\textsl{Cluster plot showing the clusters generated using the Euclidean distance in the \(K\)-means algorithm on the scaled simulated dataset.}}
        \label{fig:MahalanobisStartCluster}
    \end{minipage}
    \begin{minipage}{.45\textwidth}
        \centering
        \includegraphics[width=1\linewidth]{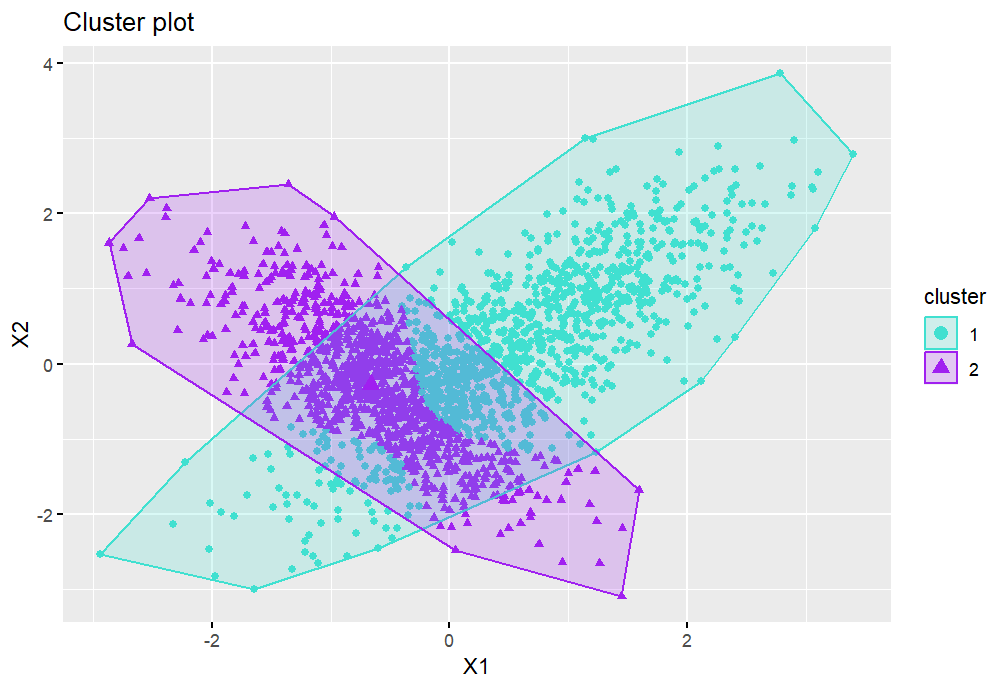}
        \caption{\textsl{Cluster plot showing the clusters produced when using the Mahalanobis distance procedure on the clusters in Figure \ref{fig:MahalanobisStartCluster}.}}
        \label{fig:MahalanobisCluster}
    \end{minipage}
\end{figure}

\begin{figure}[ht]
    \centering
    \begin{minipage}{.45\textwidth}
        \centering
        \includegraphics[width=1\linewidth]{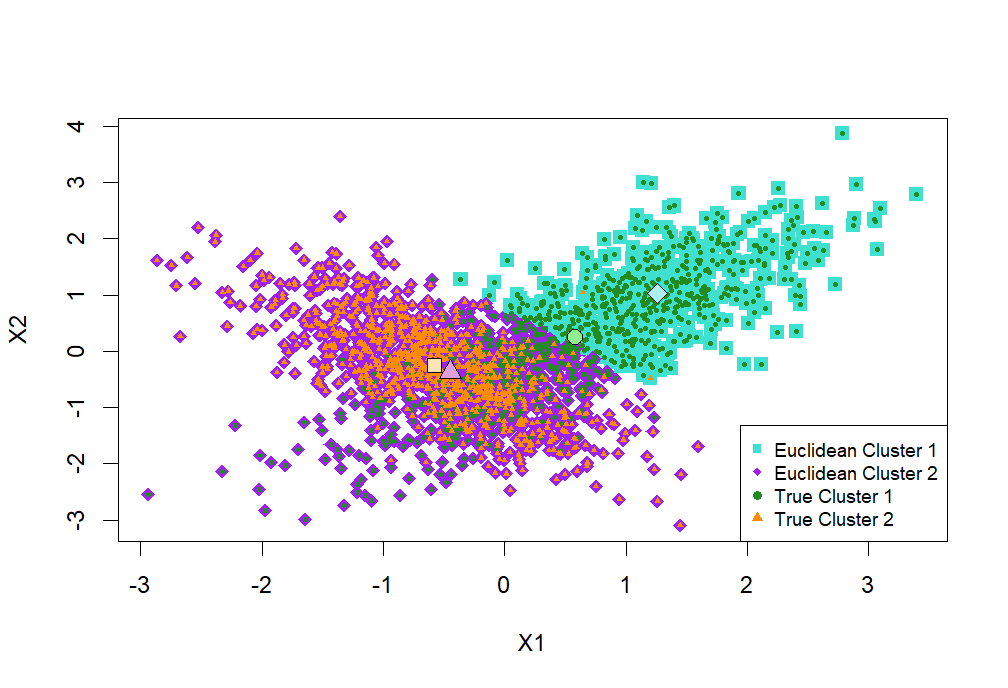}
        \caption{\textsl{Scatter plot showing the true clusters on top of the clusters generated using the Euclidean distance in the \(K\)-means algorithm on the scaled simulated dataset.}}
        \label{fig:MahalanobisStartScatter}
    \end{minipage}
    \begin{minipage}{.45\textwidth}
        \centering
        \includegraphics[width=1\linewidth]{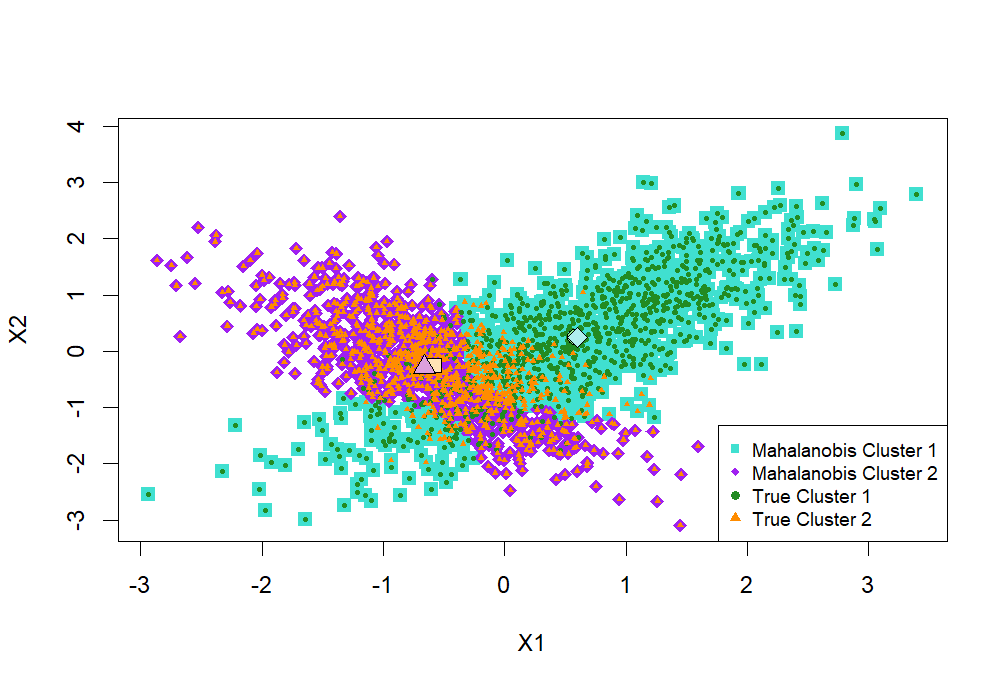}
        \caption{\textsl{Scatter plot showing the true clusters on top of the clusters when using the Mahalanobis distance procedure on the clusters in Figure \ref{fig:MahalanobisStartCluster}.}}
        \label{fig:MahalanobisScatter}
    \end{minipage}
\end{figure}

\begin{figure}[ht]
    \centering
    \begin{minipage}{.45\textwidth}
        \centering
        \includegraphics[width=1\linewidth]{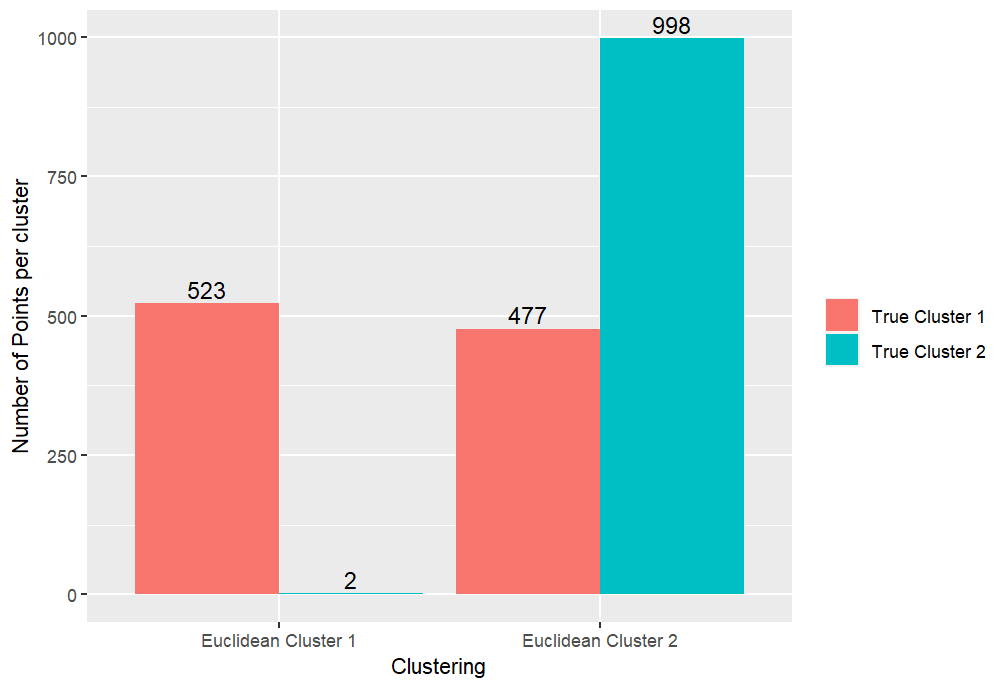}
        \caption{\textsl{Bar plot showing the number of misclassified and correctly classified points in each cluster when using the Euclidean distance in the \(K\)-means algorithm on the scaled simulated dataset.}}
        \label{fig:MahalanobisStartBar}
    \end{minipage}
    \begin{minipage}{.45\textwidth}
        \centering
        \includegraphics[width=1\linewidth]{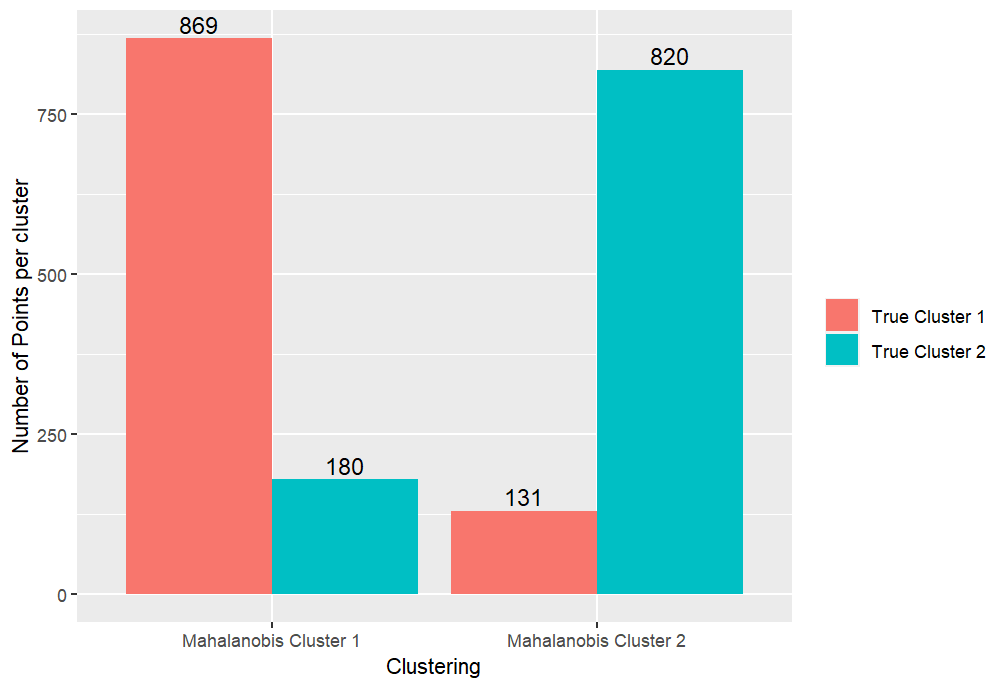}
        \caption{\textsl{Bar plot showing the number of misclassified and correctly classified points in each cluster when using the Mahalanobis distance procedure on the clusters in Figure \ref{fig:MahalanobisStartCluster}.}}
        \label{fig:MahalanobisBar}
    \end{minipage}
\end{figure}

For our final exploration on this simulated dataset, we investigate the affect of using the Mahalanobis distance in the clustering of the points. As discussed in Section \ref{subsec:KMeansClustering}, in order to use the Mahalanobis distance for clustering, we first need to to apply \(K\)-means clustering with the Euclidean distance to get an initial clustering. We can then try to improve this initial clustering by taking the covariance structure of the clusters into account via the Mahalanobis distance. We create the same graphs as for the previous sections, but this time we have two of each: one for the initial Euclidean clustering, and one for the final grouping after using the Mahalanobis distance. Figures \ref{fig:MahalanobisStartCluster} and \ref{fig:MahalanobisCluster} show respectively the initial and final clusters when using the Mahalanobis distance. We can observe that, as before, the Euclidean distance measure has resulted in two non-overlapping groups, with a much smaller cluster 1 than the true clustering. On the other hand, it would seem that, as predicted, the Mahalanobis distance appears to have worked well here. This is because we can see from Figure \ref{fig:MahalanobisCluster} that the clusters have now adopted the long intersecting shape of the true clusters. Comparing to Figure \ref{fig:TrueCluster}, we can see that the overall outline of cluster 1 seems almost exactly right, having now picked up the points from the bottom left that have been absent from the generated cluster 1 for all other distance measures. Furthermore, apart from being slightly narrower, the second generated cluster seems to approximate the true cluster shape reasonably well. The scatter plots for the initial and final scenarios can be seen in Figures \ref{fig:MahalanobisStartScatter} and \ref{fig:MahalanobisScatter}. From Figure \ref{fig:MahalanobisScatter}, we can see that the Mahalanobis cluster centroids are now almost completely aligned with the true cluster centroids. The main difference between this and the true one of Figure \ref{fig:TrueScatter} is in the intersection, where both groups share similar traits anyway. We can numerically view the change in accuracy by comparing the bar plots in Figures \ref{fig:MahalanobisStartBar} and \ref{fig:MahalanobisBar} and the Tables \ref{subtab:MahalanobisStart} and \ref{subtab:Mahalanobis}. From this, we can see that, whilst the number of points in the true group 2 misclassified has increased, the number of true group 1 points misclassified has greatly reduced. Moreover, the total number of points misclassified has reduced from 479 to 311, which is a significant improvement. In addition, the points that are misclassified following the use of the Mahalanobis distance tend to be in the overlap section, whereas when using just the Euclidean distance, the misclassified points tend to be the group 1 points in the bottom left, altering the shape of the clusters dramatically. 

\begin{table}[ht]
    \begin{subtable}[h]{0.41\textwidth}
        \centering
        \begin{adjustbox}{width=\textwidth,center}
        \begin{tabular}{|c|c|c|}
            \hline
             & Euclidean Cluster 1 & Euclidean Cluster 2\\
            \hline 
            True Cluster 1 & 523 & 477\\
            \hline
            True Cluster 1 & 2 & 998\\
            \hline
        \end{tabular}
        \end{adjustbox}
        \caption{\textsl{Tabular representation of the correctly and incorrectly classified points following the application of the \(K\)-means algorithm with the Euclidean distance on the scaled simulated dataset.}}
        \label{subtab:MahalanobisStart}
    \end{subtable}
    \hfill
    \begin{subtable}[h]{0.49\textwidth}
        \centering
        \begin{adjustbox}{width=\textwidth,center}
        \begin{tabular}{|c|c|c|}
            \hline
             & Mahalanobis Cluster 1 & Mahalanobis Cluster 2\\
            \hline 
            True Cluster 1 & 869 & 131\\
            \hline
            True Cluster 1 & 180 & 820\\
            \hline
        \end{tabular}
        \end{adjustbox}
        \caption{\textsl{Tabular representation of the correctly and incorrectly classified points when using the Mahalanobis distance procedure on the clusters in Figure \ref{fig:MahalanobisStartCluster}.}}
        \label{subtab:Mahalanobis}
    \end{subtable}
    \caption{\textsl{Comparison of the number of points correctly and incorrectly classified before and after using the Mahalanobis distance to update the clustering.}}
\end{table}

\newpage
In summary, for this simulated dataset specifically, we may conclude that the Mahalanobis distance has been highly effective at reducing the number of points misclassified following the clustering and producing the correct overall shape of the clusters. Beyond this, the other distance measures perform relatively similarly, with the next best distance measures of those considered being the Maximum and the Euclidean. The Manhattan distance appears to perform the worst here. In terms of the Maximum and Euclidean distances, the better performance in terms of number of misclassified is the Maximum distance by a very marginal amount, whilst the Euclidean distance seems to perform better in terms of the overall shape of the clusters. However, we have seen that this can be very dependent on the underlying dataset. Thus, given more time, to improve this work further, we should consider additional simulated data with varying properties and potentially higher dimensions, and compare the results found in these cases in order to gain a deeper understanding of this topic. In the next section, we move on to repeating these experiments on subsets of the \emph{Dry Bean} dataset to see if similar results can be found for this case. 

\subsection{The \emph{Dry Bean} Dataset}\label{subsec:DatasetExample}
In this section, we consider the \emph{Dry Bean} dataset, which is available from the UCI Machine Learning Repository \cite{misc_dry_bean_dataset_602}. Here, however, we restrict ourselves to the six variables \emph{Area}, \emph{Perimeter}, \emph{MajorAxisLength}, \emph{ConvexArea}, \emph{EquivDiameter} and \emph{ShapeFactor3} that showed strong linear relations in our investigations from the previous coursework. In addition, we filter the dataset to consider two separate cases of just two classes of dry bean.

\subsubsection{Example 1}
\begin{figure}[ht]
    \centering
    \begin{minipage}{.45\textwidth}
        \centering
        \includegraphics[width=1\linewidth]{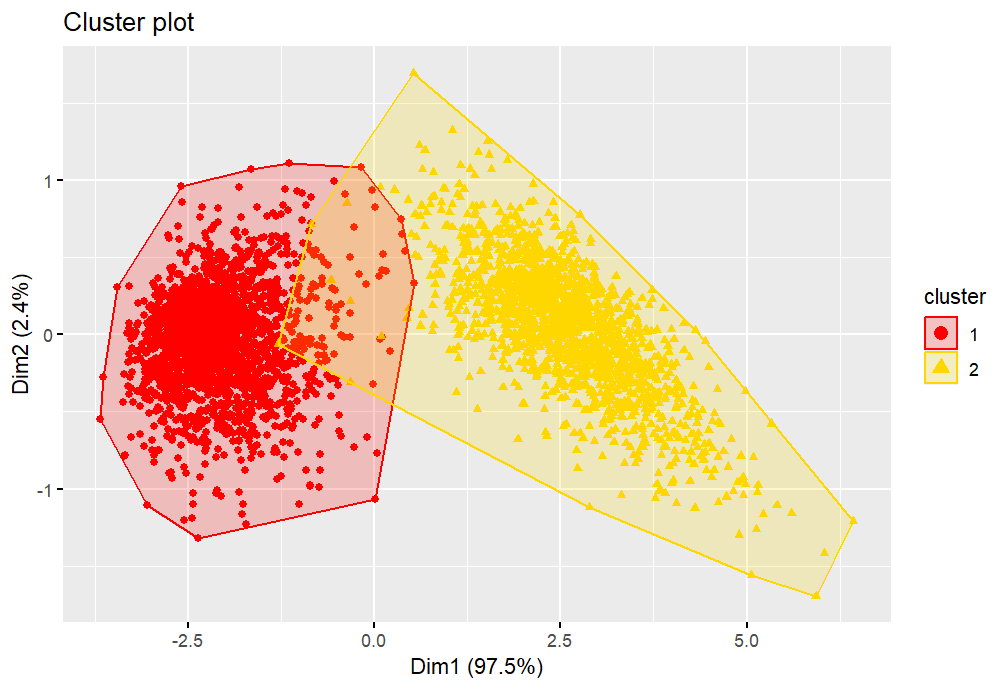}
        \caption{\textsl{Cluster plot when using the true clusters for this subset of the \emph{Dry Bean} dataset.}}
        \label{fig:TrueClusterPlotData1}
    \end{minipage}
    \begin{minipage}{.45\textwidth}
        \centering
        \includegraphics[width=1\linewidth]{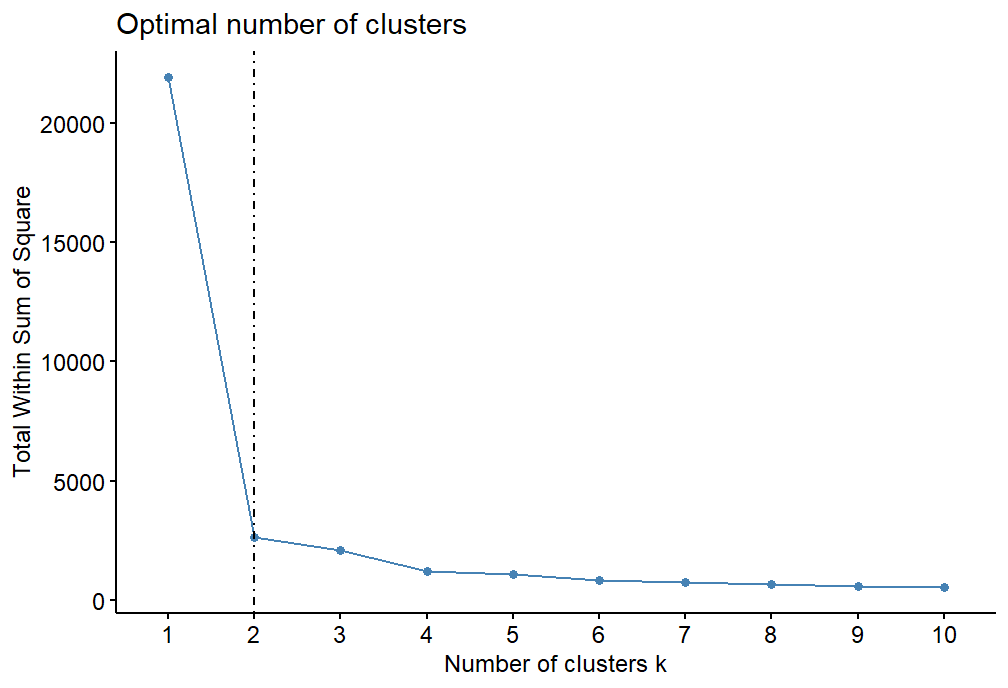}
        \caption{\textsl{Scree plot for this subset of the \emph{Dry Bean} dataset.}}
        \label{fig:ScreePlotData1}
    \end{minipage}
\end{figure}

Here, we consider the classes \emph{SEKER} and \emph{CALI}, though from this point forward we will simply refer to them as groups 1 and 2 respectively. We again perform the necessary scaling to enable us to carry out the clustering, and we note that the first two principal components are used in order to graphically represent the results. Figure \ref{fig:TrueClusterPlotData1} shows what the cluster plot should look like given the correct clustering. We observe that the cumulative percentage variance explained for the first two dimensions is 99.9\%, so that the two-dimensional plots we display are good representations of the data. We can notice that the two clusters are fairly well separated, with one being highly spherical as well. This could imply that the Euclidean distance will work well. The scree plot of Figure \ref{fig:ScreePlotData1} has a clear elbow at two, suggesting that we are correct in selecting \(K=2\) in the \(K\)-means algorithm.

\begin{figure}[ht]
    \centering
    \begin{minipage}{.45\textwidth}
        \centering
        \includegraphics[width=1\linewidth]{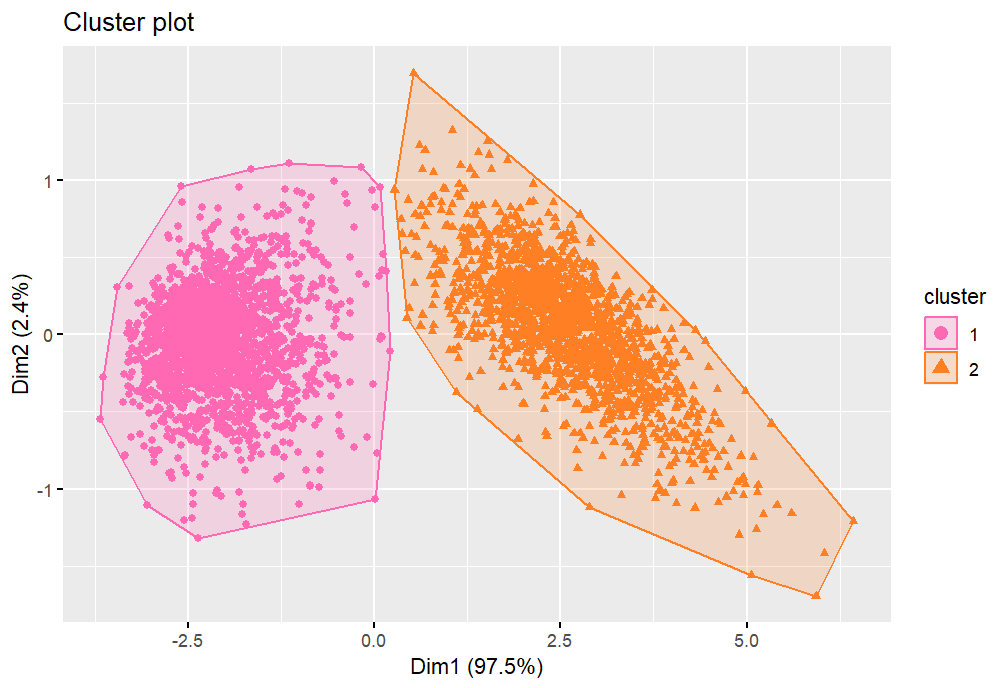}
        \caption{\textsl{Cluster plot showing the clusters produced following the application of the \(K\)-means algorithm with the Euclidean distance on this subset of the \emph{Dry Bean} dataset.}}
        \label{fig:EuclideanClusterPlotData1}
    \end{minipage}
    \begin{minipage}{.45\textwidth}
        \centering
        \includegraphics[width=1\linewidth]{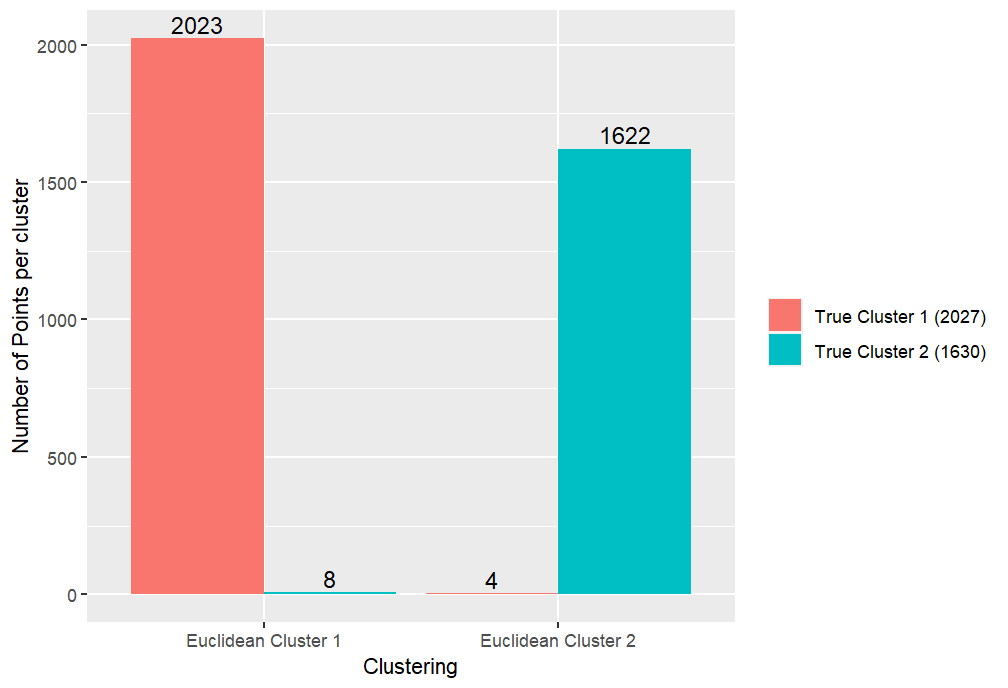}
        \caption{\textsl{Bar plot showing the number of misclassified and correctly classified points in each cluster when using the Euclidean distance in the \(K\)-means algorithm on this subset of the \emph{Dry Bean} dataset.}}
        \label{fig:EuclideanBarPlotData1}
    \end{minipage}
\end{figure}

\begin{figure}[p!]
    \centering
    \begin{minipage}{.45\textwidth}
        \centering
        \includegraphics[width=1\linewidth]{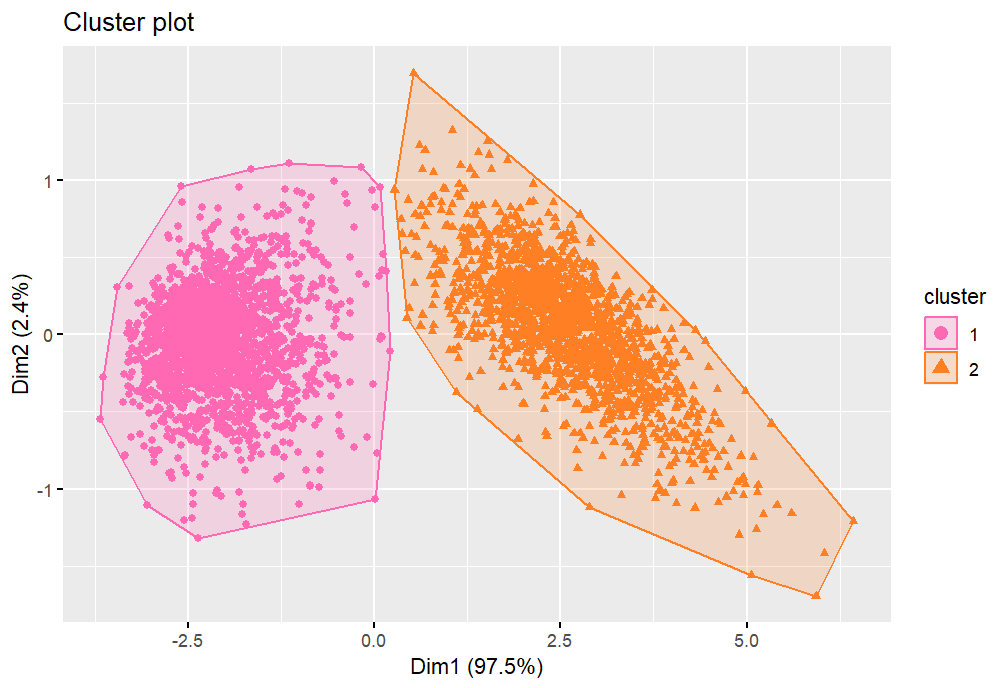}
        \caption{\textsl{Cluster plot showing the clusters produced following the application of the \(K\)-means algorithm with the Manhattan distance on this subset of the \emph{Dry Bean} dataset.}}
        \label{fig:ManhattanClusterPlotData1}
    \end{minipage}
    \begin{minipage}{.45\textwidth}
        \centering
        \includegraphics[width=1\linewidth]{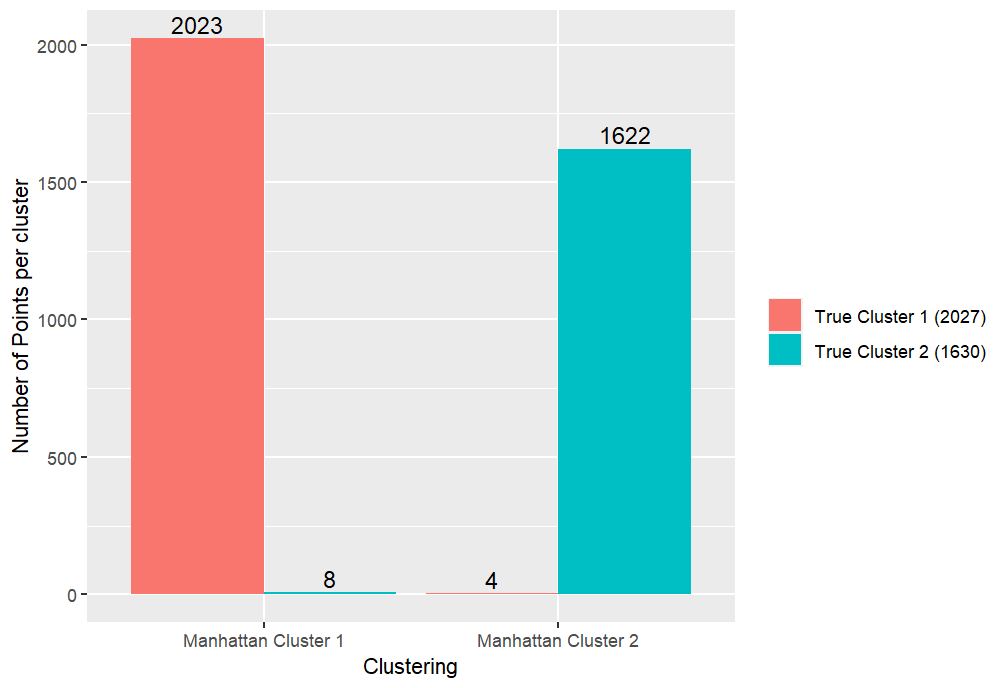}
        \caption{\textsl{Bar plot showing the number of misclassified and correctly classified points in each cluster when using the Manhattan distance in the \(K\)-means algorithm on this subset of the \emph{Dry Bean} dataset.}}
        \label{fig:ManhattanBarPlotData1}
    \end{minipage}
\end{figure}

\begin{figure}[p!]
    \centering
    \begin{minipage}{.45\textwidth}
        \centering
        \includegraphics[width=1\linewidth]{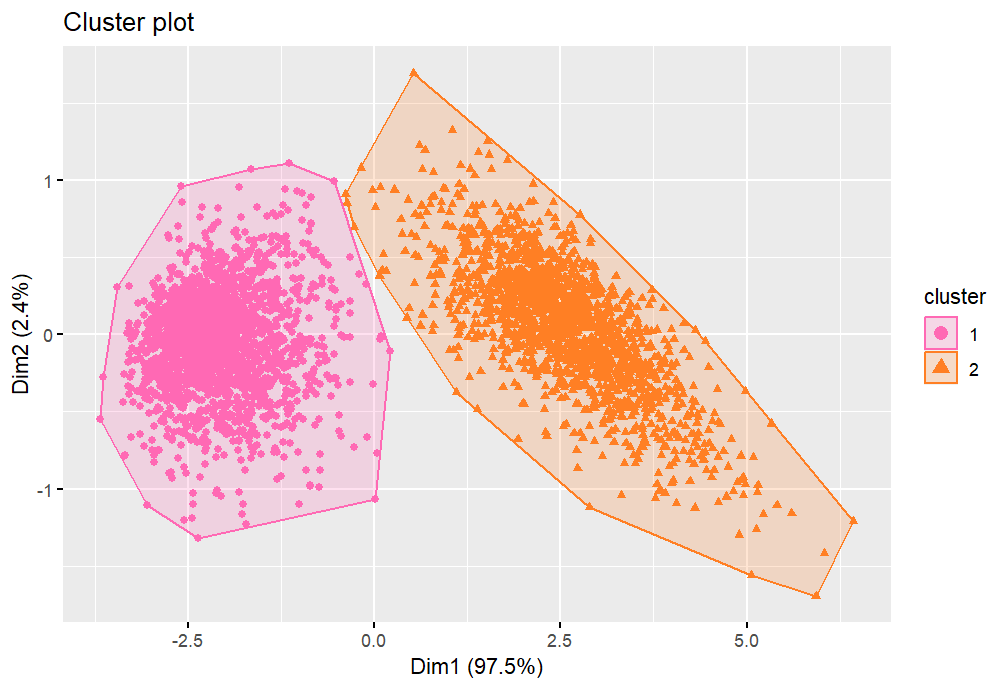}
        \caption{\textsl{Cluster plot showing the clusters produced following the application of the \(K\)-means algorithm with the Maximum distance on this subset of the \emph{Dry Bean} dataset.}}
        \label{fig:MaximumClusterPlotData1}
    \end{minipage}
    \begin{minipage}{.45\textwidth}
        \centering
        \includegraphics[width=1\linewidth]{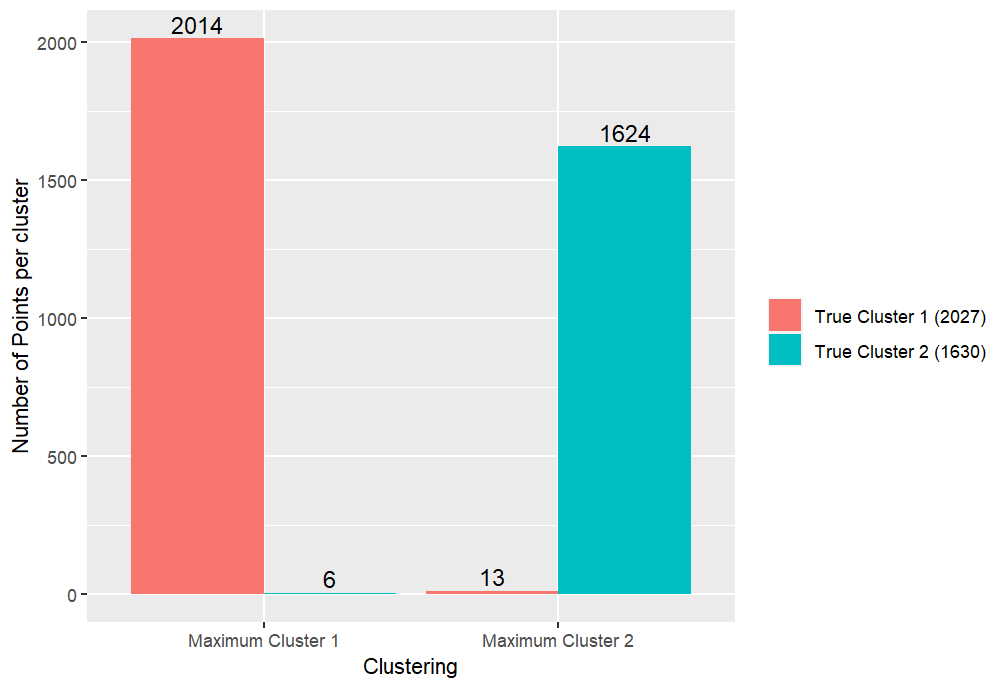}
        \caption{\textsl{Bar plot showing the number of misclassified and correctly classified points in each cluster when using the Maximum distance in the \(K\)-means algorithm on this subset of the \emph{Dry Bean} dataset.}}
        \label{fig:MaximumBarPlotData1}
    \end{minipage}
\end{figure}

Now, Figures \ref{fig:EuclideanClusterPlotData1}, \ref{fig:ManhattanClusterPlotData1} and \ref{fig:MaximumClusterPlotData1} show the cluster plots for the Euclidean, Manhattan and Maximum distance measures respectively. Likewise Figures \ref{fig:EuclideanBarPlotData1}, \ref{fig:ManhattanBarPlotData1} and \ref{fig:MaximumBarPlotData1} contain the corresponding bar plots of correctly and incorrectly classified data points. For the data we use here, we see that the Euclidean and Manhattan distances give us the same clustering, where we remark that both appear to be very good. This is because there are only 12 of the 3657 points misclassified. This could be because the underlying data is much more separated than in Section~\ref{subsec:SimulatedExample}. However, we can notice that it seems to be the points in the overlap that are misclassified. In particular the general shape of the first cluster appears to be highly accurate and it is just the shape of the second in the top right that is incorrect. The Maximum distance seems to capture the shape of the second cluster slightly better, however it performs marginally worse in terms of the number of points misclassified. Overall though, these three metrics give very similar outcomes for this specific scenario.

\begin{figure}[p!]
    \centering
    \begin{minipage}{.45\textwidth}
        \centering
        \includegraphics[width=1\linewidth]{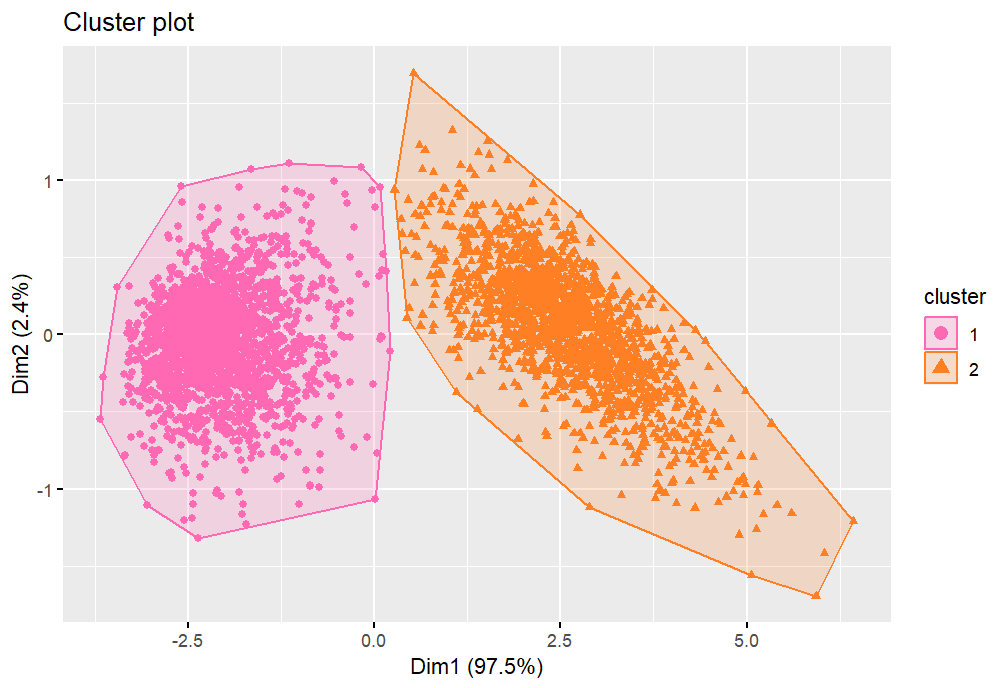}
        \caption{\textsl{Cluster plot showing the clusters produced following the application of the \(K\)-means algorithm with the Euclidean distance on this subset of the \emph{Dry Bean} dataset.}}
        \label{fig:MahalanobisStartClusterData1}
    \end{minipage}
    \begin{minipage}{.45\textwidth}
        \centering
        \includegraphics[width=1\linewidth]{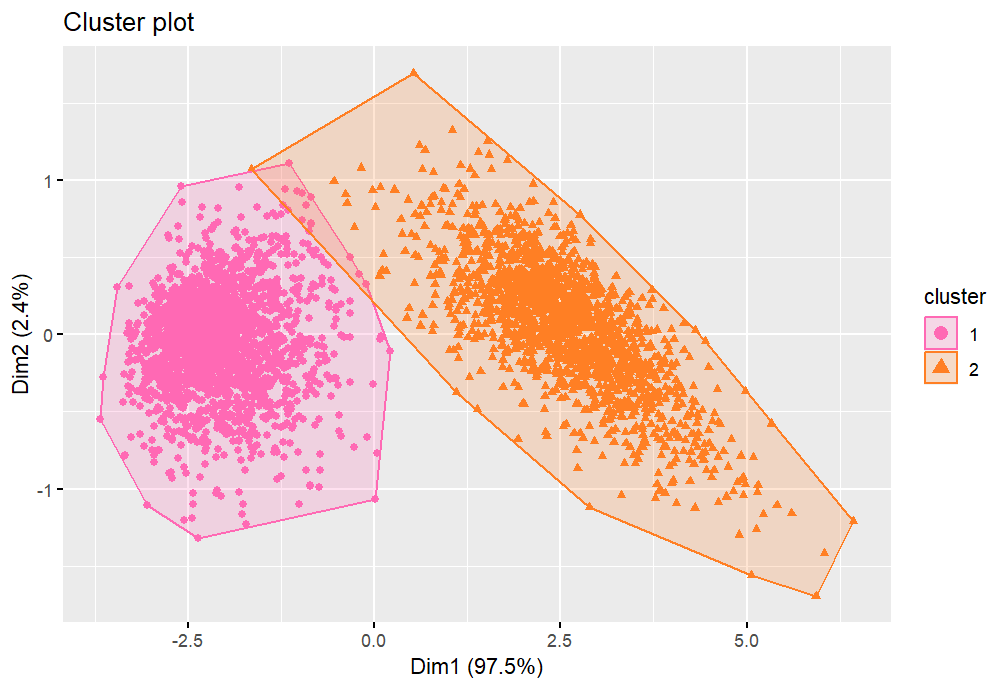}
        \caption{\textsl{Cluster plot showing the clusters produced when using the Mahalanobis distance procedure on the clusters in Figure \ref{fig:MahalanobisStartClusterData1}.}}
        \label{fig:MahalanobisClusterPlotData1}
    \end{minipage}
\end{figure}

\begin{figure}[htp!]
    \centering
    \begin{minipage}{.45\textwidth}
        \centering
        \includegraphics[width=1\linewidth]{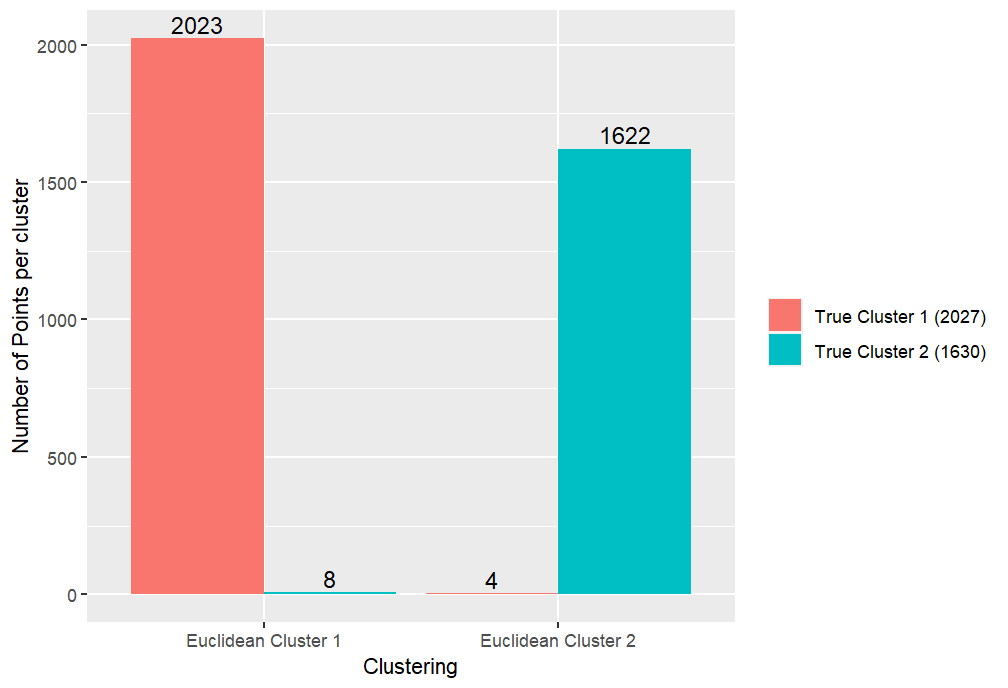}
        \caption{\textsl{Bar plot showing the number of misclassified and correctly classified points in each cluster when using the Euclidean distance in the \(K\)-means algorithm on the scaled simulated dataset.}}
        \label{fig:MahalanobisStartBarData1}
    \end{minipage}
    \begin{minipage}{.45\textwidth}
        \centering
        \includegraphics[width=1\linewidth]{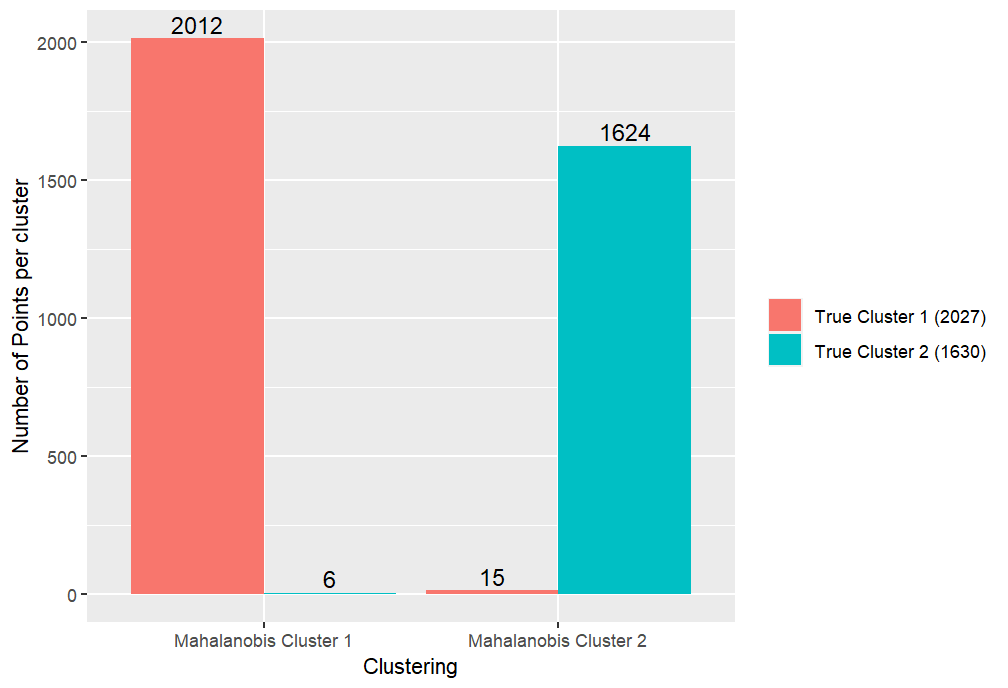}
        \caption{\textsl{Bar plot showing the number of misclassified and correctly classified points in each cluster when using the Mahalanobis distance procedure on the clusters in Figure \ref{fig:MahalanobisStartClusterData1}.}}
        \label{fig:MahalanobisBarData1}
    \end{minipage}
\end{figure}

The cluster plots for before and after using the Mahalanobis distance can be seen in Figures \ref{fig:MahalanobisStartClusterData1} and \ref{fig:MahalanobisClusterPlotData1}. Likewise, the corresponding bar plots can be seen in Figures \ref{fig:MahalanobisStartBarData1} and \ref{fig:MahalanobisBarData1}. When comparing between the initial clusters generated using the Euclidean distance and the final result after using the Mahalanobis distance we can see that there is much less difference visible in this case than for our simulated example of Section \ref{subsec:SimulatedExample}. In terms of recognising the slight overlap of the clusters, the Mahalanobis distance may have been useful. By contrast though, the total number of points misclassified actually increases slightly after using the Mahalanobis distance. This suggests that in this scenario, the Mahalanobis distance seems to have neither greatly hindered nor drastically improved on the results obtained using the Euclidean distance. In the next section, we repeat the experiments we have performed here using a different pair of clusters. 

\newpage
\subsubsection{Example 2}
\begin{figure}[ht]
    \centering
    \begin{minipage}{.45\textwidth}
        \centering
        \includegraphics[width=1\linewidth]{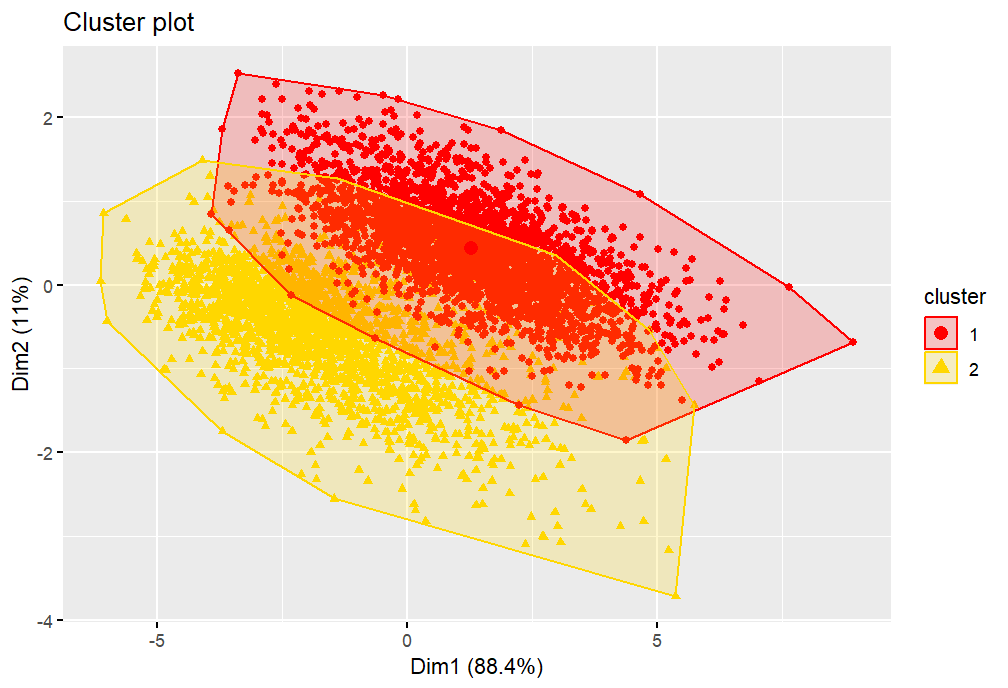}
        \caption{\textsl{Cluster plot when using the true clusters for this subset of the \emph{Dry Bean} dataset.}}
        \label{fig:TrueClusterPlotData2}
    \end{minipage}
    \begin{minipage}{.45\textwidth}
        \centering
        \includegraphics[width=1\linewidth]{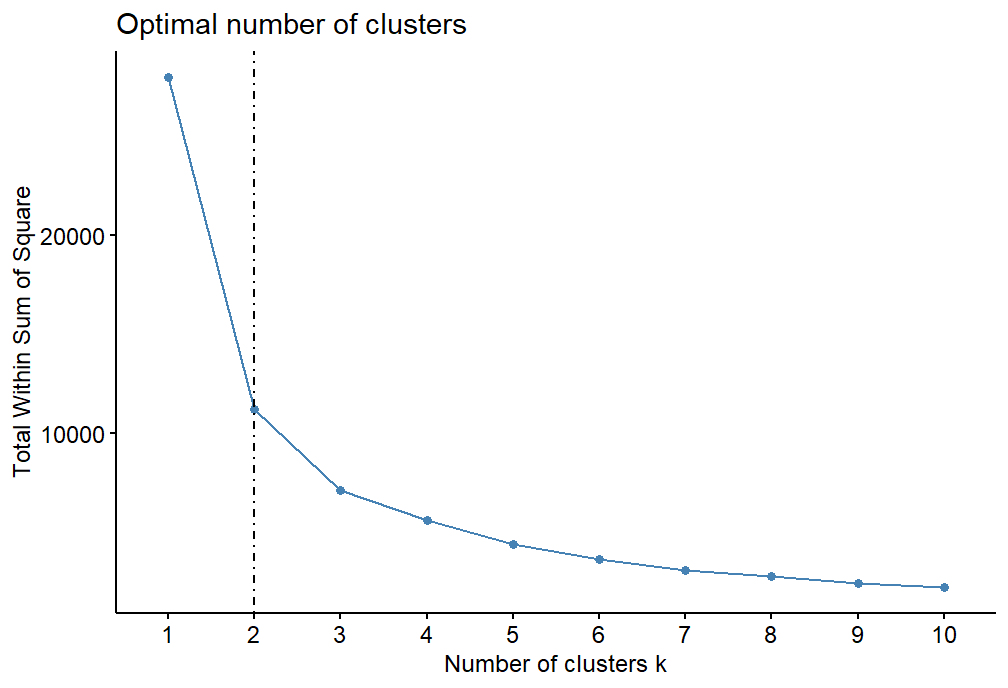}
        \caption{\textsl{Scree plot for this subset of the \emph{Dry Bean} dataset.}}
        \label{fig:ScreePlotData2}
    \end{minipage}
\end{figure}

As our final example we consider the classes \emph{SIRA} and \emph{SEKER}, though as before, we simply consider them to be the true clusters 1 and 2 respectively. Looking at the scatter plot of Figure \ref{fig:TrueClusterPlotData2}, we can see that this time the two clusters are much more similar in shape and location. Note that the clusters appear elliptical, but contrary to the simulated example of Section \ref{subsec:SimulatedExample}, both clusters are in the same direction. This may make it more difficult for the covariance structure to distinguish between them. We again give the scree plot for this data (Figure \ref{fig:ScreePlotData2}). Though not quite as sharp as in Figure \ref{fig:ScreePlotData1}, it would still seem apparent that \(K=2\) is the sensible option for the number of clusters. This suggests that, despite the similarity between the clusters, the \(K\)-means algorithm has been able to identify that there should be two groups. 

\begin{figure}[ht]
    \centering
    \begin{minipage}{.45\textwidth}
        \centering
        \includegraphics[width=1\linewidth]{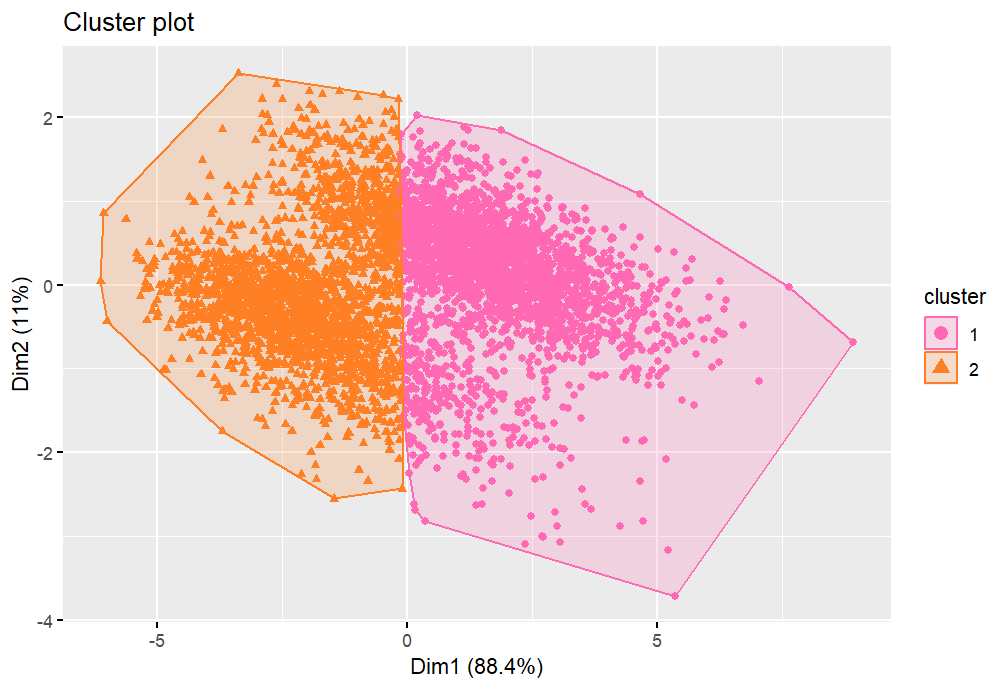}
        \caption{\textsl{Cluster plot showing the clusters produced following the application of the \(K\)-means algorithm with the Euclidean distance on this subset of the \emph{Dry Bean} dataset.}}
        \label{fig:EuclideanClusterPlotData2}
    \end{minipage}
    \begin{minipage}{.45\textwidth}
        \centering
        \includegraphics[width=1\linewidth]{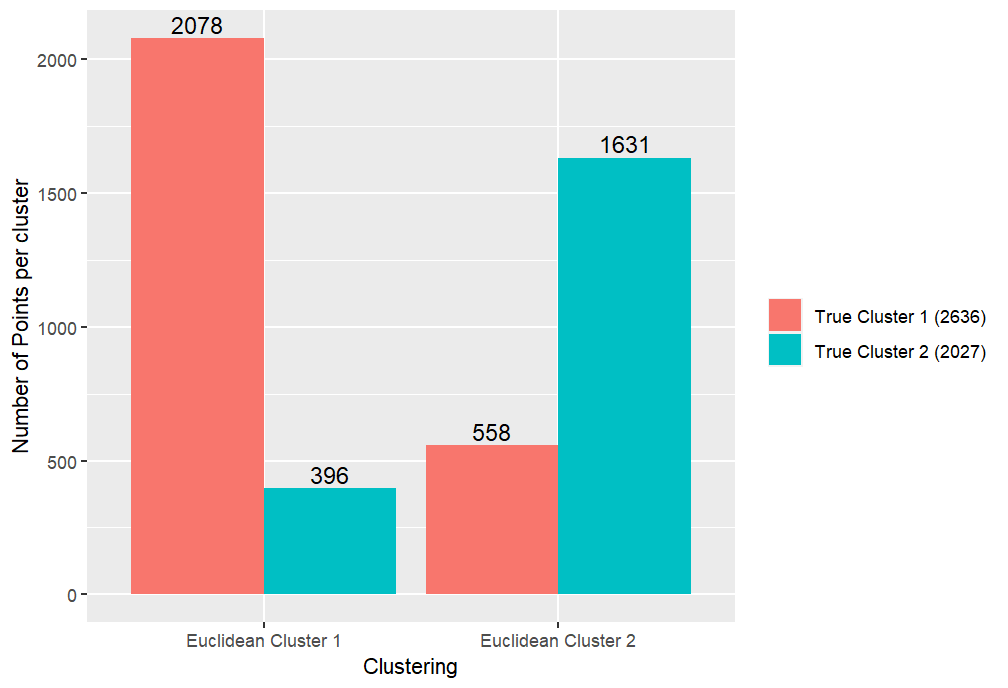}
        \caption{\textsl{Bar plot showing the number of misclassified and correctly classified points in each cluster when using the Euclidean distance in the \(K\)-means algorithm on this subset of the \emph{Dry Bean} dataset.}}
        \label{fig:EuclideanBarPlotData2}
    \end{minipage}
\end{figure}

\begin{figure}[ht]
    \centering
    \begin{minipage}{.45\textwidth}
        \centering
        \includegraphics[width=1\linewidth]{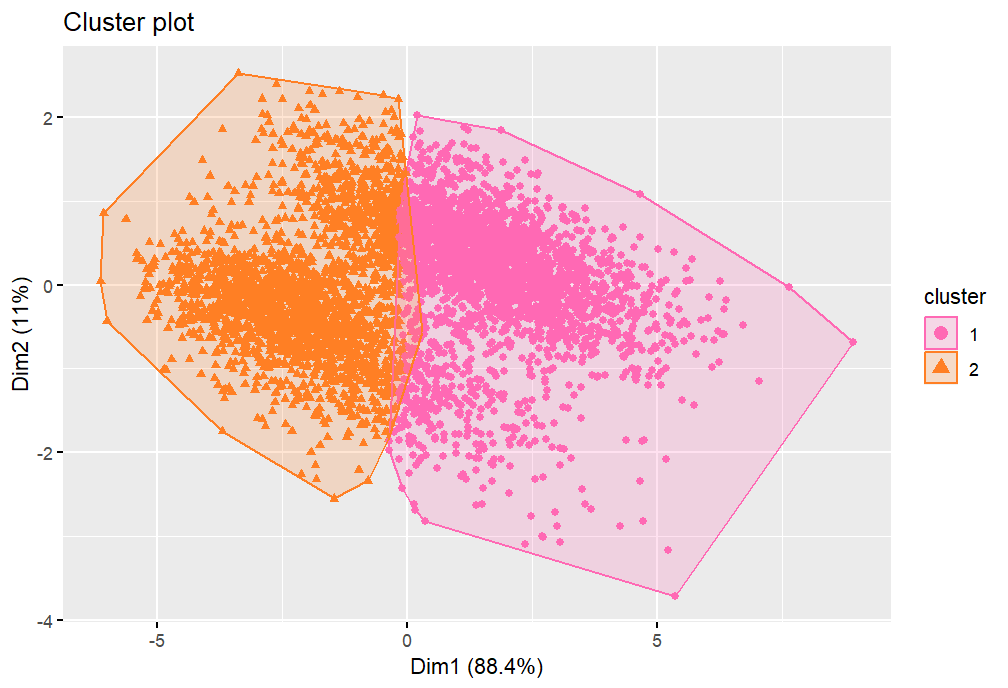}
        \caption{\textsl{Cluster plot showing the clusters produced following the application of the \(K\)-means algorithm with the Manhattan distance on this subset of the \emph{Dry Bean} dataset.}}
        \label{fig:ManhattanClusterPlotData2}
    \end{minipage}
    \begin{minipage}{.45\textwidth}
        \centering
        \includegraphics[width=1\linewidth]{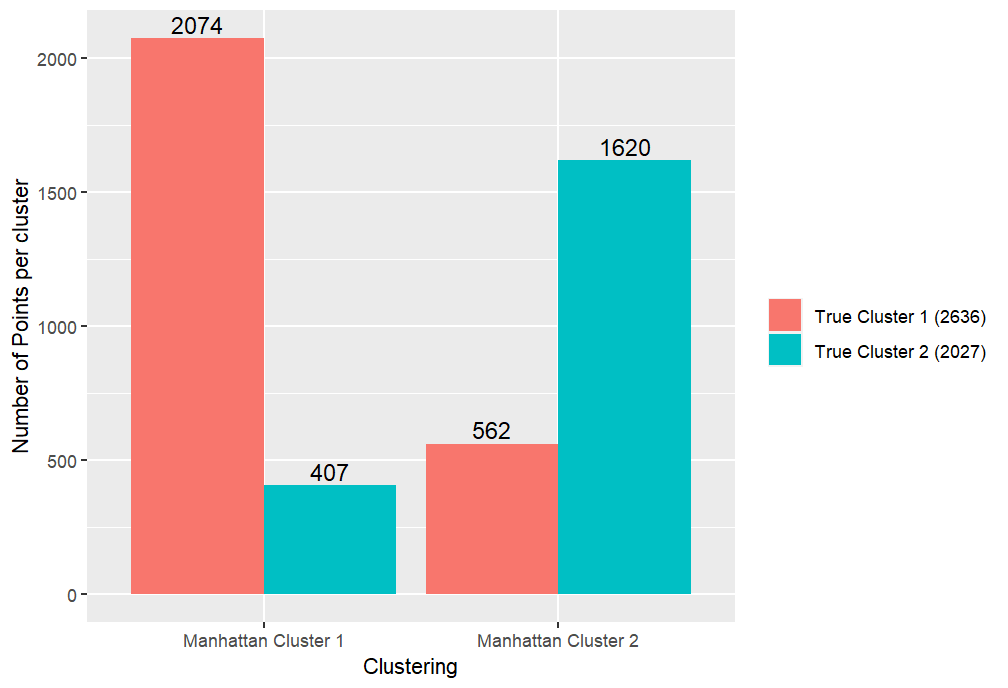}
        \caption{\textsl{Bar plot showing the number of misclassified and correctly classified points in each cluster when using the Manhattan distance in the \(K\)-means algorithm on this subset of the \emph{Dry Bean} dataset.}}
        \label{fig:ManhattanBarPlotData2}
    \end{minipage}
\end{figure}

\begin{figure}[hpt]
    \centering
    \begin{minipage}{.45\textwidth}
        \centering
        \includegraphics[width=1\linewidth]{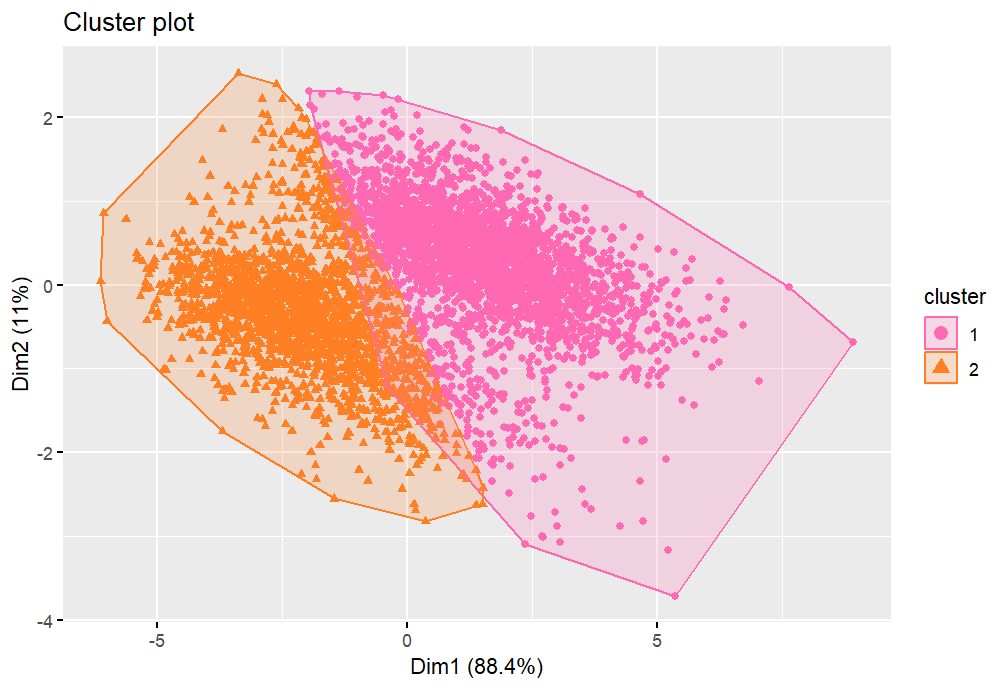}
        \caption{\textsl{Cluster plot showing the clusters produced following the application of the \(K\)-means algorithm with the Maximum distance on this subset of the \emph{Dry Bean} dataset.}}
        \label{fig:MaximumClusterPlotData2}
    \end{minipage}
    \begin{minipage}{.45\textwidth}
        \centering
        \includegraphics[width=1\linewidth]{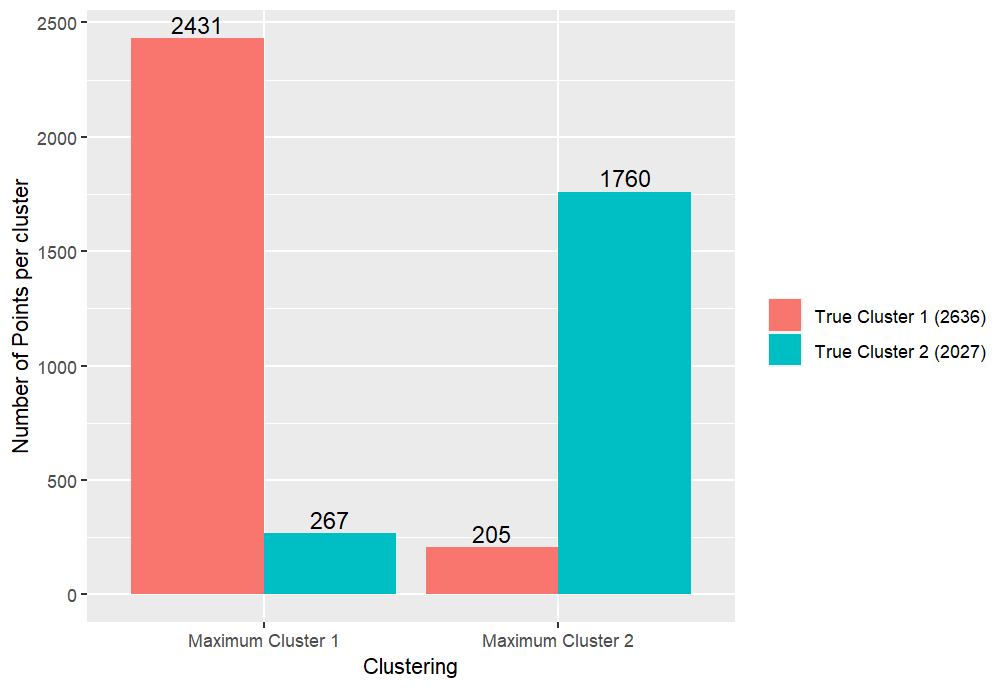}
        \caption{\textsl{Bar plot showing the number of misclassified and correctly classified points in each cluster when using the Maximum distance in the \(K\)-means algorithm on this subset of the \emph{Dry Bean} dataset.}}
        \label{fig:MaximumBarPlotData2}
    \end{minipage}
\end{figure}

We now compare Figures \ref{fig:EuclideanClusterPlotData2}, \ref{fig:ManhattanClusterPlotData2} and \ref{fig:MaximumClusterPlotData2} which contain the cluster plots resulting from the Euclidean, Manhattan and Maximum distances respectively, as previously. In addition, we observe the bar plots comparing the misclassification errors that can be seen in Figures \ref{fig:EuclideanBarPlotData2}, \ref{fig:ManhattanBarPlotData2} and \ref{fig:MaximumBarPlotData2}, respectively. Here, we observe that none of the distance measures appear to have found the desired clustering. The Euclidean and Manhattan distances again seem to have found similar clusterings, though they have quite a high number of misclassified points due to their vertical partitioning of the data points. By contrast, the Maximum distance appears to have performed better here, with half the number of incorrectly classified points compared to the other two metrics. This seems to be because it has segmented the data more on a diagonal and so seems to encapsulate more of the underlying structure of the clusters.

\begin{figure}[hpt]
    \centering
    \begin{minipage}{.45\textwidth}
        \centering
        \includegraphics[width=1\linewidth]{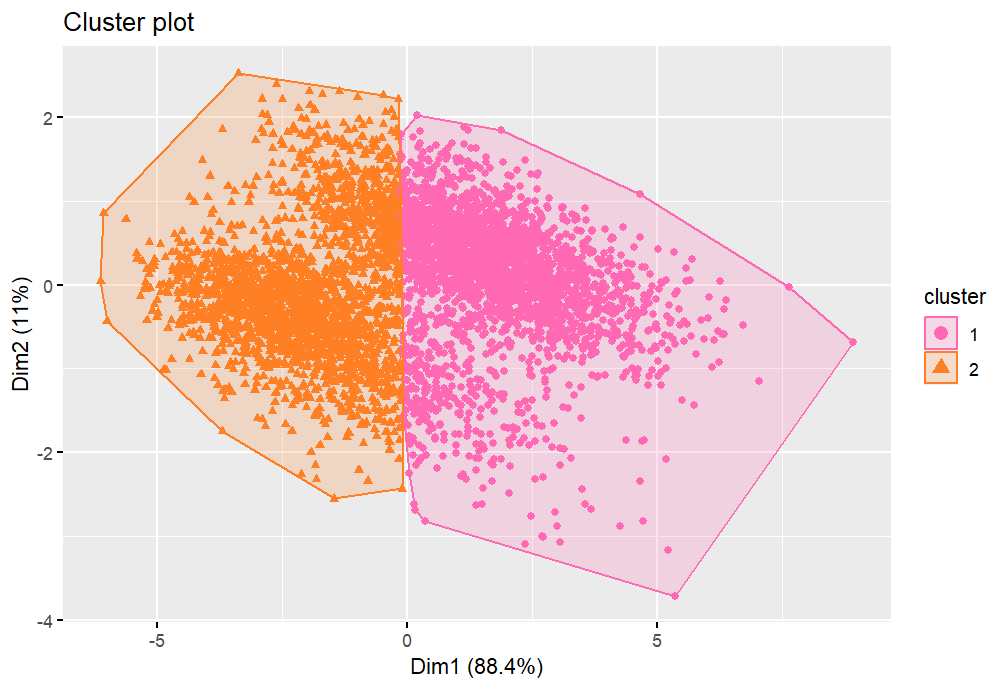}
        \caption{\textsl{Cluster plot showing the clusters produced following the application of the \(K\)-means algorithm with the Euclidean distance on this subset of the \emph{Dry Bean} dataset.}}
        \label{fig:MahalanobisStartClusterData2}
    \end{minipage}
    \begin{minipage}{.45\textwidth}
        \centering
        \includegraphics[width=1\linewidth]{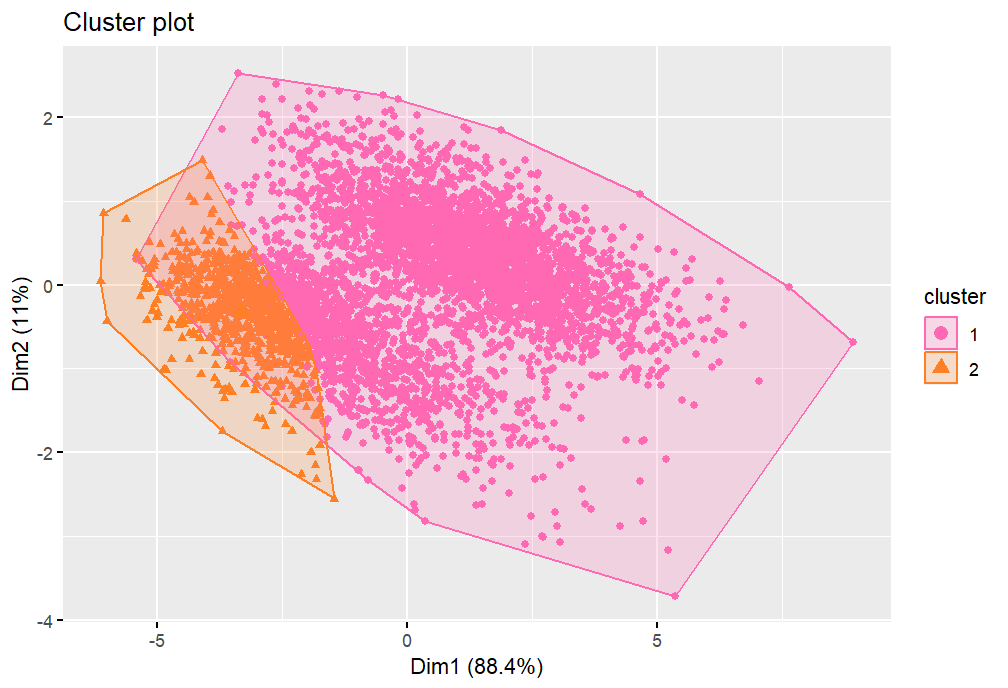}
        \caption{\textsl{Cluster plot showing the clusters produced when using the Mahalanobis distance procedure on the clusters in Figure \ref{fig:MahalanobisStartClusterData2}.}}
        \label{fig:MahalanobisClusterPlotData2}
    \end{minipage}
\end{figure}

\begin{figure}[hpt!]
    \centering
    \begin{minipage}{.45\textwidth}
        \centering
        \includegraphics[width=1\linewidth]{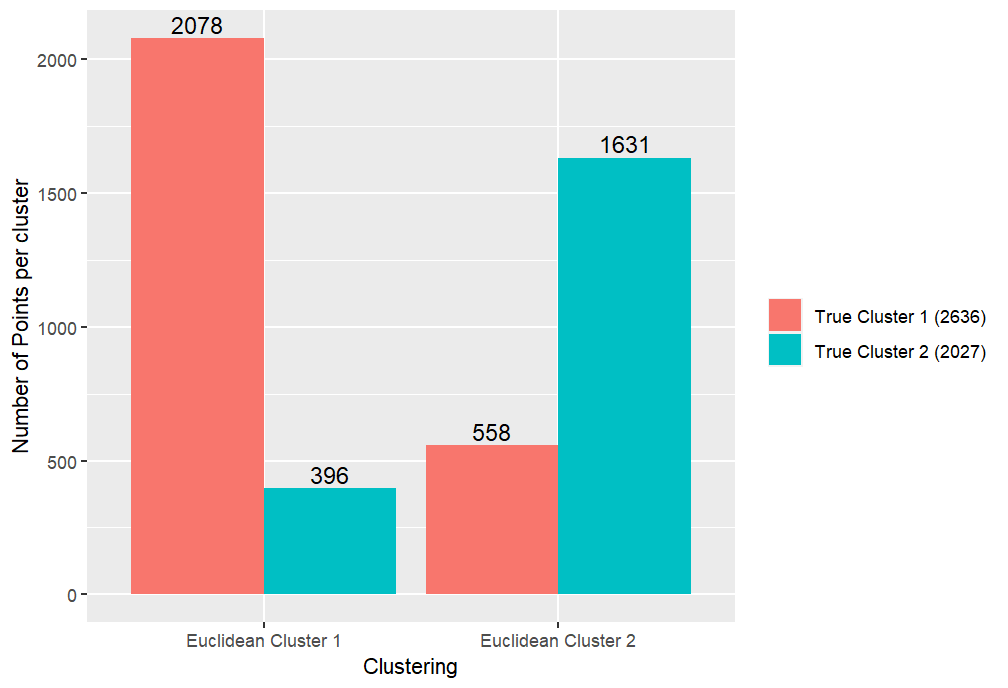}
        \caption{\textsl{Bar plot showing the number of misclassified and correctly classified points in each cluster when using the Euclidean distance in the \(K\)-means algorithm on the scaled simulated dataset.}}
        \label{fig:MahalanobisStartBarData2}
    \end{minipage}
    \begin{minipage}{.45\textwidth}
        \centering
        \includegraphics[width=1\linewidth]{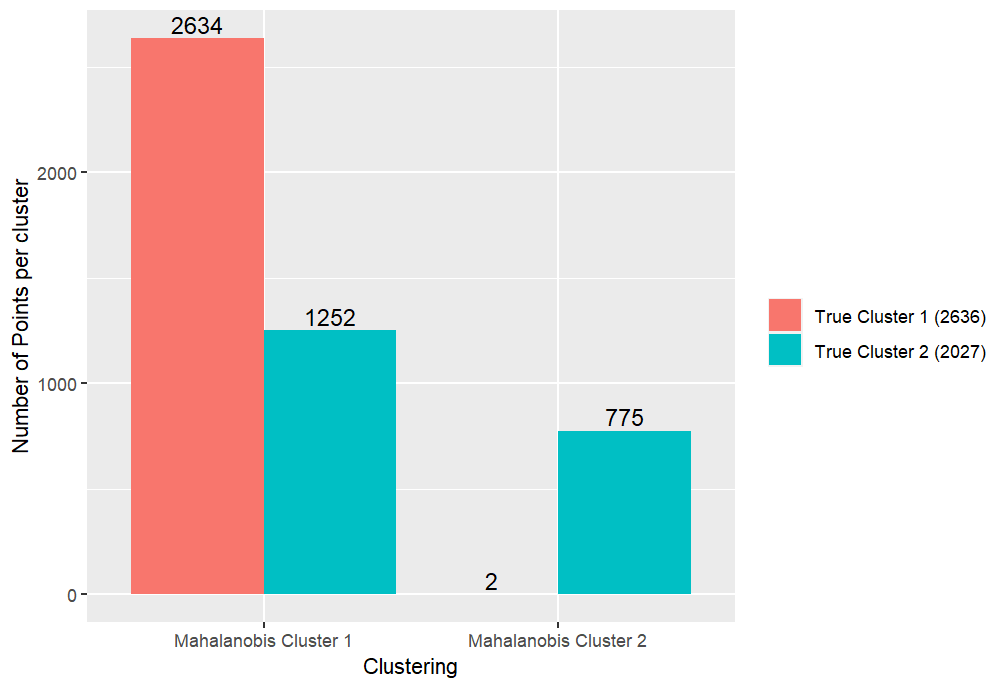}
        \caption{\textsl{Bar plot showing the number of misclassified and correctly classified points in each cluster when using the Mahalanobis distance procedure on the clusters in Figure \ref{fig:MahalanobisStartClusterData2}.}}
        \label{fig:MahalanobisBarData2}
    \end{minipage}
\end{figure}

Next, we use the Mahalanobis distance on a clustering produced using the Euclidean distance in the \(K\)-means algorithm. The cluster plots and bar plots for before and after the application of the Mahalanobis distance can be seen in Figures \ref{fig:MahalanobisStartClusterData2}, \ref{fig:MahalanobisClusterPlotData2}, \ref{fig:MahalanobisStartBarData2} and \ref{fig:MahalanobisBarData2}. This time, we may observe that the Mahalanobis distance seems to have worsened the clustering given by the Euclidean distance, by putting the majority of points into the one cluster. This could suggest that perhaps there is greater similarity between the covariance matrices of the two clusters, resulting in the Mahalanobis distance being less effective in this case. 

In summary, for the two subsets of the \emph{Dry Bean} dataset we have considered, we have seen the importance of the underlying structure of the data in the performance quality of the different distance measures discussed. Whilst the Mahalanobis distance has been seen to be effective in the simulated example of Section \ref{subsec:SimulatedExample}, it has appears to have had little or worsening effect on our examples in this section. Furthermore, whilst we had previously seen that the Euclidean distance tended to be at least as good at the Manhattan and Maximum distances, in our second example we have seen a scenario in which the Maximum distance appears to be the superior method in terms of the total number of points misclassified. This again emphasises the importance of understanding the structure of the underlying dataset and trialling out different methods to compare results before selecting a distance measure, since there is no one best solution in all cases. 

\newpage
\section{Conclusion}\label{sec:Conclusion}
In conclusion, this report has investigated the effect of using different distance measures in the \(K\)-means clustering algorithm. This has been achieved by exploring literature relating to this topic and by applying the distance measures to some real-life and simulated datasets. 

Firstly, in Section \ref{subsec:Papers}, we explored some papers linked to distance measures for clustering. In general, we found the overarching opinion that the inherent characteristics of the dataset to be clustered are very influential on the performance of the distance measures, since using different datasets can lead to different conclusions of the best distance measure to use. Additionally, we have seen that there are scenarios where the Mahalanobis distance is implemented in practice, but we have also seen situations where the more traditional Euclidean distance is the preferred metric. Furthermore, there are also generalisations to the Mahalanobis distance that exist and can appear to work well in certain scenarios. However, these do not seem to have been either further researched or implemented. We could develop this section further in the future by considering additional papers, particularly some more recent ones, to see if there has been a shift in the attitude towards the effectiveness of the Mahalanobis distance. 

Next, in Section \ref{subsec:ChatGPT} we investigated the opinion of ChatGPT in this area. We found that, whilst it appears to have some general knowledge in this area, it is not able to provide specific answers to questions. In addition, there are occasions where we may question its reliability and accuracy. To improve this section further, we could consider asking some more detailed questions, such as by giving it a specific dataset to consider, to see if this can produce more direct answers. We could also repeat the asked questions a number of times to assess its reproducibility and reliability. 

Finally, in Section \ref{sec:Applications}, we applied the Euclidean, Manhattan, Maximum and Mahalanobis distance measures in the \(K\)-means clustering algorithm on a simulated dataset and two subsets of the \emph{Dry Bean} dataset. We found that the Mahalanobis distance gave extremely good improvements over the other distances for the simulated dataset, being able to identify the intersecting, highly elliptical clusters well. However, the Mahalanobis distance gave underwhelming performances for the other two examples tried. In these situations, the Euclidean/Manhattan and Maximum distance measures respectively gave the best results. This demonstrates the importance of considering the underlying structure and characteristics of the dataset when deciding on a distance measure. It also serves to suggest the importance of trialling several methods to compare the results before selecting a distance measure since there is no evidence to suggest that one distance measure will always outperform another. Given more time, to improve this investigation further, we should consider implementing the different distance measures on many more example datasets both from real-life and simulated, to attempt to understand in what scenarios the different distance measures tend to work well.

\subsection*{Acknowledgement}
The author is very grateful to Professor Anatoly Zhigljavsky for attracting the author's attention to this problem and for providing useful advice.

\newpage
\section{References}
\printbibliography[heading=none]

\newpage
\appendix
\section{Appendix}\label{sec:Appendix}
In this section, we present the R code used to perform the clustering and produce the graphs presented throughout Section \ref{sec:Applications}. Specifically, in Section \ref{subsec:SimulatedDatasetCode} we give the code used to create the simulated dataset and that for all the experiments we performed on it in Section \ref{subsec:SimulatedExample}. Then, in Section \ref{subsec:DatasetCode}, we give the code used for the \emph{Dry Bean} dataset examples explored in Section \ref{subsec:DatasetExample}. 

\subsection{Code for the Simulated Dataset (Section \ref{subsec:SimulatedExample})}\label{subsec:SimulatedDatasetCode}
\begin{lstlisting}[breaklines=true, language=R, commentstyle=\color{teal}]
# K-Means Cluster Analysis on a 2D Simulated Data Set.

rm(list=ls())

# Load necessary packages.
library(MASS)
library(factoextra)
library(amap)
library(ggplot2)

# Set the seed for reproducibility of results.
set.seed(4511)

# Simulate 1000 instances of two different bivariate normal distributions.
# Covariance matrices need to be symmetric positive semi-definite - check using eigen()
# Should get non-negative eigenvalues.
mu1 <- c(1,1)
mu2 <- c(-0.5,0.5)
sigma1 <- matrix(c(1.5,1,1,1),2,2)
sigma2 <- matrix(c(0.8,-0.5,-0.5,0.6),2,2)
all(eigen(sigma1)$values > 0)
all(eigen(sigma2)$values > 0)
first_data_points <- mvrnorm(1000,mu=mu1,Sigma=sigma1)
second_data_points <- mvrnorm(1000,mu=mu2,Sigma=sigma2)
classifier1 <- rep(1,1000)
classifier2 <- rep(2,1000)
colnames(first_data_points) <- c("X1", "X2")
colnames(second_data_points) <- c("X1", "X2")

# Create the simulated data frame and true clusters.
df_original <- data.frame(rbind(first_data_points,second_data_points))
true.clusters <- c(classifier1,classifier2)
colours <- c("forestgreen","darkorange")
point_types <- c(16,17)

# Scale the data.
df <- scale(df_original)
true_centre_1 <- colMeans(df[1:1000,])
true_centre_2 <- colMeans(df[1001:2000,])

# Plot the scaled data.
plot(df[,1],df[,2],col=colours[factor(true.clusters)],
     pch=point_types[factor(true.clusters)],
     xlab="X1",ylab="X2")
points(true_centre_1[1],true_centre_1[2],bg="lightgreen",pch=21,col="black",cex=1.5)
points(true_centre_2[1],true_centre_2[2],bg="moccasin",pch=24,col="black",cex=1.5)
legend("bottomright",legend=c("True Cluster 1", "True Cluster 2"),
       pch=point_types,col=colours,cex=0.8)

# View the covariance matrix of the scaled data.
cov(df) 

# Find the true clusters and true cluster centres.
true.clusters <-c(classifier1,classifier2)
true.centers <- rbind(true_centre_1,true_centre_2)
colnames(true.centers) <- c("X1", "X2")

# View the appearance of the cluster plot given the correct clustering.
fviz_cluster(object=list(data=df,cluster=true.clusters),palette=colours,repel=TRUE,
             geom="point",show.clust.cent=TRUE,ellipse=TRUE,ellipse.type="convex")

##########################################################################
##########################################################################

# Use the K-Means algorithm with Euclidean distance to classify the points.
res.Euclid <- Kmeans(df,2,iter.max=100,nstart=100,method="euclidean")
Euclidean.clusters <- res.Euclid$cluster
Euclidean.centers <- res.Euclid$centers

colours2 <- c("turquoise","purple")
point_types2 <- c(15,18)

# Cluster plot from Euclidean distance.
fviz_cluster(object=list(data=df,cluster=Euclidean.clusters),palette=colours2,repel=TRUE,
             geom="point",show.clust.cent=TRUE,ellipse=TRUE,ellipse.type="convex")

# Scatter plot from Euclidean distance.
plot(df[,1],df[,2],col=colours2[factor(Euclidean.clusters)],
     pch=point_types2[factor(Euclidean.clusters)],cex=1.3,
     xlab="X1",ylab="X2")
points(df[,1],df[,2],col=colours[factor(true.clusters)],
       pch=point_types[factor(true.clusters)],cex=0.5)
points(true_centre_1[1],true_centre_1[2],bg="lightgreen",pch=21,col="black",cex=1.5)
points(true_centre_2[1],true_centre_2[2],bg="moccasin",pch=22,col="black",cex=1.5)
points(Euclidean.centers[1,1],Euclidean.centers[1,2],bg="paleturquoise",pch=23,
       col="black",cex=1.5)
points(Euclidean.centers[2,1],Euclidean.centers[2,2],bg="plum",pch=24,
       col="black",cex=1.5)
legend("bottomright",legend=c("Euclidean Cluster 1", "Euclidean Cluster 2",
                              "True Cluster 1", "True Cluster 2"),
       pch=c(point_types2,point_types),col=c(colours2,colours),cex=0.8)

# Create clustered bar plot of missclassified results. 
# Code inspired by https://r-graph-gallery.com/48-grouped-barplot-with-ggplot2.
cluster_1_table <- table(Euclidean.clusters[1:1000])
cluster_2_table <- table(Euclidean.clusters[1001:2000])
clustering <- c(rep("Euclidean Cluster 1",2), rep("Euclidean Cluster 2",2))
true_clustering <- rep(c("True Cluster 1" , "True Cluster 2"), 2)
values <- c(cluster_1_table[1],cluster_2_table[1],cluster_1_table[2],cluster_2_table[2])
data <- data.frame(clustering,true_clustering,values)

ggplot(data, aes(fill=true_clustering, y=values, x=clustering)) + 
  geom_bar(position="dodge", stat="identity") +
  xlab("Clustering") +
  ylab("Number of Points per cluster") +
  geom_col(position = "dodge") +
  geom_text(
    aes(label = values),
    colour = "black", size = 4,
    vjust = -0.3, position = position_dodge(.9)
  ) +
  labs(fill=" ") 

# Tabulate the number of points correctly and incorrectly classified into clusters 1
# and 2 from Euclidean distance.
table(true.clusters,Euclidean.clusters)

##########################################################################
##########################################################################

# Use the K-Means algorithm with Manhattan distance to classify the points.
set.seed(2024)
res.Manhattan <- Kmeans(df,2,iter.max=100,nstart=100,method="manhattan")
Manhattan.clusters <- res.Manhattan$cluster
Manhattan.centers <- res.Manhattan$centers

# Cluster plot from Manhattan distance.
fviz_cluster(object=list(data=df,cluster=Manhattan.clusters),palette=colours2,repel=TRUE,
             geom="point",show.clust.cent=TRUE,ellipse=TRUE,ellipse.type="convex")

# Scatter plot from Manhattan distance.
plot(df[,1],df[,2],col=colours2[factor(Manhattan.clusters)],
     pch=point_types2[factor(Manhattan.clusters)],cex=1.3,
     xlab="X1",ylab="X2")
points(df[,1],df[,2],col=colours[factor(true.clusters)],
       pch=point_types[factor(true.clusters)],cex=0.5)
points(true_centre_1[1],true_centre_1[2],bg="lightgreen",pch=21,col="black",cex=1.5)
points(true_centre_2[1],true_centre_2[2],bg="moccasin",pch=22,col="black",cex=1.5)
points(Manhattan.centers[1,1],Manhattan.centers[1,2],bg="paleturquoise",pch=24,
       col="black",cex=1.5)
points(Manhattan.centers[2,1],Manhattan.centers[2,2],bg="plum",pch=23,
       col="black",cex=1.5)
legend("bottomright",legend=c("Manhattan Cluster 1", "Manhattan Cluster 2",
                              "True Cluster 1", "True Cluster 2"),
       pch=c(point_types2,point_types),col=c(rev(colours2),colours),cex=0.8)

# Create clustered bar plot of missclassified results. 
# Code inspired by https://r-graph-gallery.com/48-grouped-barplot-with-ggplot2.
cluster_1_table <- table(Manhattan.clusters[1:1000])
cluster_2_table <- table(Manhattan.clusters[1001:2000])
clustering <- c(rep("Manhattan Cluster 1",2), rep("Manhattan Cluster 2",2))
true_clustering <- rep(c("True Cluster 1" , "True Cluster 2"), 2)
values <- c(cluster_1_table[1],cluster_2_table[1],cluster_1_table[2],cluster_2_table[2])
data <- data.frame(clustering,true_clustering,values)

ggplot(data, aes(fill=true_clustering, y=values, x=clustering)) + 
  geom_bar(position="dodge", stat="identity") +
  xlab("Clustering") +
  ylab("Number of Points per cluster") +
  geom_col(position = "dodge") +
  geom_text(
    aes(label = values),
    colour = "black", size = 4,
    vjust = -0.3, position = position_dodge(.9)
  ) +
  labs(fill=" ") 

# Tabulate the number of points correctly and incorrectly classified into clusters 1
# and 2 from Manhattan distance.
table(true.clusters,Manhattan.clusters)

##########################################################################
##########################################################################

# Use the K-Means algorithm with Maximum distance to classify the points.
set.seed(1234)
res.max <- Kmeans(df,2,iter.max=100,nstart=100,method="maximum")
max.clusters <- res.max$cluster
max.centers <- res.max$centers

# Cluster plot from Maximum distance.
fviz_cluster(object=list(data=df,cluster=max.clusters),palette=colours2,repel=TRUE,
             geom="point",show.clust.cent=TRUE,ellipse=TRUE,ellipse.type="convex")

# Scatter plot from Maximum distance.
plot(df[,1],df[,2],col=colours2[factor(max.clusters)],
     pch=point_types2[factor(max.clusters)],cex=1.3,
     xlab="X1",ylab="X2")
points(df[,1],df[,2],col=colours[factor(true.clusters)],
       pch=point_types[factor(true.clusters)],cex=0.5)
points(true_centre_1[1],true_centre_1[2],bg="lightgreen",pch=21,col="black",cex=1.5)
points(true_centre_2[1],true_centre_2[2],bg="moccasin",pch=22,col="black",cex=1.5)
points(max.centers[1,1],max.centers[1,2],bg="paleturquoise",pch=23,col="black",cex=1.5)
points(max.centers[2,1],max.centers[2,2],bg="plum",pch=24,col="black",cex=1.5)
legend("bottomright",legend=c("Maximum Cluster 1", "Maximum Cluster 2",
                              "True Cluster 1", "True Cluster 2"),
       pch=c(point_types2,point_types),col=c(colours2,colours),cex=0.8)

# Create clustered bar plot of missclassified results. 
# Code inspired by https://r-graph-gallery.com/48-grouped-barplot-with-ggplot2.
cluster_1_table <- table(max.clusters[1:1000])
cluster_2_table <- table(max.clusters[1001:2000])
clustering <- c(rep("Maximum Cluster 1",2), rep("Maximum Cluster 2",2))
true_clustering <- rep(c("True Cluster 1" , "True Cluster 2"), 2)
values <- c(cluster_1_table[1],cluster_2_table[1],cluster_1_table[2],cluster_2_table[2])
data <- data.frame(clustering,true_clustering,values)

ggplot(data, aes(fill=true_clustering, y=values, x=clustering)) + 
  geom_bar(position="dodge", stat="identity") +
  xlab("Clustering") +
  ylab("Number of Points per cluster") +
  geom_col(position = "dodge") +
  geom_text(
    aes(label = values),
    colour = "black", size = 4,
    vjust = -0.3, position = position_dodge(.9)
  ) +
  labs(fill=" ") 

# Tabulate the number of points correctly and incorrectly classified into clusters 1
# and 2 from Maximum distance.
table(true.clusters,max.clusters)

##########################################################################
##########################################################################

# Use the K-Means algorithm with Mahalanobis distance to classify the points.
# Do 50 iterations with Euclidean distance first to find initial clustering.
set.seed(1234)
res.Mahalanobis <- Kmeans(df,2,iter.max=50,nstart=50,method="euclidean")
Mahalanobis.clusters <- res.Mahalanobis$cluster
Mahalanobis.centers <- res.Mahalanobis$centers
cluster1_inds <- which(Mahalanobis.clusters==1)
cluster2_inds <- which(Mahalanobis.clusters==2)
centroid1 <- Mahalanobis.centers[1,]
centroid2 <- Mahalanobis.centers[2,]
cov1 <- cov(df[cluster1_inds,])
cov2 <- cov(df[cluster2_inds,])

# Cluster plot of starting point for the Mahalanobis distance.
fviz_cluster(object=list(data=df,cluster=Mahalanobis.clusters),palette=colours2,
             repel=TRUE,geom="point",show.clust.cent=TRUE,ellipse=TRUE,
             ellipse.type="convex")

# Scatter plot of starting point for the Mahalanobis distance.
plot(df[,1],df[,2],col=colours2[factor(Mahalanobis.clusters)],
     pch=point_types2[factor(Mahalanobis.clusters)],cex=1.3,
     xlab="X1",ylab="X2")
points(df[,1],df[,2],col=colours[factor(true.clusters)],
       pch=point_types[factor(true.clusters)],cex=0.5)
points(true_centre_1[1],true_centre_1[2],bg="lightgreen",pch=21,col="black",cex=1.5)
points(true_centre_2[1],true_centre_2[2],bg="moccasin",pch=22,col="black",cex=1.5)
points(centroid1[1],centroid1[2],bg="paleturquoise",pch=23,col="black",cex=1.5)
points(centroid2[1],centroid2[2],bg="plum",pch=24,col="black",cex=1.5)
legend("bottomright",legend=c("Euclidean Cluster 1", "Euclidean Cluster 2",
                              "True Cluster 1", "True Cluster 2"),
       pch=c(point_types2,point_types),col=c(colours2,colours),cex=0.8)

# Create clustered bar plot of missclassified results for starting point of Mahalanobis
# distance. Code inspired by https://r-graph-gallery.com/48-grouped-barplot-with-ggplot2.
cluster_1_table <- table(Mahalanobis.clusters[1:1000])
cluster_2_table <- table(Mahalanobis.clusters[1001:2000])
clustering <- c(rep("Euclidean Cluster 1",2), rep("Euclidean Cluster 2",2))
true_clustering <- rep(c("True Cluster 1" , "True Cluster 2"), 2)
values <- c(cluster_1_table[1],cluster_2_table[1],cluster_1_table[2],cluster_2_table[2])
data <- data.frame(clustering,true_clustering,values)

ggplot(data, aes(fill=true_clustering, y=values, x=clustering)) + 
  geom_bar(position="dodge", stat="identity") +
  xlab("Clustering") +
  ylab("Number of Points per cluster") +
  geom_col(position = "dodge") +
  geom_text(
    aes(label = values),
    colour = "black", size = 4,
    vjust = -0.3, position = position_dodge(.9)
  ) +
  labs(fill=" ") 

# Tabulate the number of points correctly and incorrectly classified into clusters 1
# and 2 for the starting point for the Mahalanobis distance.
table(true.clusters,Mahalanobis.clusters)

# Perform the Mahalanobis distance part of the algorithm.
for (i in 1:100){
  Mahalanobis.clusters <- ifelse(
    mahalanobis(df,centroid1,cov1) < mahalanobis(df,centroid2,cov2), 
    1,
    2
  )
  cluster1_inds <- which(Mahalanobis.clusters==1)
  cluster2_inds <- which(Mahalanobis.clusters==2)
  centroid1 <- colMeans(df[cluster1_inds,])
  centroid2 <- colMeans(df[cluster2_inds,])
  cov1 <- cov(df[cluster1_inds,])
  cov2 <- cov(df[cluster2_inds,])
}

# Cluster plot from Mahalanobis distance.
fviz_cluster(object=list(data=df,cluster=Mahalanobis.clusters),palette=colours2,
             repel=TRUE,geom="point",show.clust.cent=TRUE,ellipse=TRUE,
             ellipse.type="convex")

# Scatter plot from Mahalanobis distance.
plot(df[,1],df[,2],col=colours2[factor(Mahalanobis.clusters)],
     pch=point_types2[factor(Mahalanobis.clusters)],cex=1.3,
     xlab="X1",ylab="X2")
points(df[,1],df[,2],col=colours[factor(true.clusters)],
       pch=point_types[factor(true.clusters)],cex=0.5)
points(true_centre_1[1],true_centre_1[2],bg="lightgreen",pch=21,col="black",cex=1.5)
points(true_centre_2[1],true_centre_2[2],bg="moccasin",pch=22,col="black",cex=1.5)
points(centroid1[1],centroid1[2],bg="paleturquoise",pch=23,col="black",cex=1.5)
points(centroid2[1],centroid2[2],bg="plum",pch=24,col="black",cex=1.5)
legend("bottomright",legend=c("Mahalanobis Cluster 1", "Mahalanobis Cluster 2",
                              "True Cluster 1", "True Cluster 2"),
       pch=c(point_types2,point_types),col=c(colours2,colours),cex=0.8)

# Create clustered bar plot of missclassified results. 
# Code inspired by https://r-graph-gallery.com/48-grouped-barplot-with-ggplot2.
cluster_1_table <- table(Mahalanobis.clusters[1:1000])
cluster_2_table <- table(Mahalanobis.clusters[1001:2000])
clustering <- c(rep("Mahalanobis Cluster 1",2), rep("Mahalanobis Cluster 2",2))
true_clustering <- rep(c("True Cluster 1" , "True Cluster 2"), 2)
values <- c(cluster_1_table[1],cluster_2_table[1],cluster_1_table[2],cluster_2_table[2])
data <- data.frame(clustering,true_clustering,values)

ggplot(data, aes(fill=true_clustering, y=values, x=clustering)) + 
  geom_bar(position="dodge", stat="identity") +
  xlab("Clustering") +
  ylab("Number of Points per cluster") +
  geom_col(position = "dodge") +
  geom_text(
    aes(label = values),
    colour = "black", size = 4,
    vjust = -0.3, position = position_dodge(.9)
  ) +
  labs(fill=" ") 

# Tabulate the number of points correctly and incorrectly classified into clusters 1
# and 2 from Mahalanobis distance.
table(true.clusters,Mahalanobis.clusters)

##########################################################################
##########################################################################

# Decide on the number of clusters.
fviz_nbclust(df, kmeans, method = "wss") +
  geom_vline(xintercept = 2, linetype = 4)
\end{lstlisting}

\subsection{Code for the \emph{Dry Bean} Dataset (Section \ref{subsec:DatasetExample})}\label{subsec:DatasetCode}
\begin{lstlisting}[breaklines=true, language=R, commentstyle=\color{teal}]
# Load necessary packages.
library(factoextra)
library(readxl)
library(MASS)
library(amap)
library(ggplot2)

# Import the Dry Bean Dataset.
dry_bean_data <- read_xlsx('Data/Dry_Bean_Dataset.xlsx')

# Correct two column names.
colnames(dry_bean_data)[colnames(dry_bean_data) == 'AspectRation'] <- 'AspectRatio'
colnames(dry_bean_data)[colnames(dry_bean_data) == 'roundness'] <- 'Roundness'

# Check for missing values.
colSums(is.na(dry_bean_data))

# Create a subset of the whole dataset.
dataset <- dry_bean_data[,c(1,2,3,7,8,15,17)]

##########################################################################
##########################################################################

# Choose two classes to keep for the first investigation.
inds1 <- which(dataset[,7]=="CALI"|dataset[,7]=="SEKER")
data1 <- dataset[inds1,1:6]
group1 <- dataset[inds1,7]
group1 <- ifelse(group1=="SEKER",1,2)
table(group1)

# Set colour palette to colour the different Class types by.
colours <- c("red","gold")

# Scale the data.
df <- scale(data1)

# View the appearance of the cluster plot given the correct clustering.
fviz_cluster(object=list(data=df,cluster=group1),palette=colours,repel=TRUE,
             geom="point",show.clust.cent=TRUE,ellipse=TRUE,ellipse.type="convex")

##########################################################################
##########################################################################

# Use the K-Means algorithm with Euclidean distance to classify the points.
set.seed(4511)
res.Euclid <- Kmeans(df,2,iter.max=100,nstart=100,method="euclidean")
Euclidean.clusters <- res.Euclid$cluster
Euclidean.centers <- res.Euclid$centers

colours2 <- c("hotpink","chocolate1")

# Cluster plot from Euclidean distance.
fviz_cluster(object=list(data=df,cluster=Euclidean.clusters),palette=colours2,repel=TRUE,
             geom="point",show.clust.cent=TRUE,ellipse=TRUE,ellipse.type="convex")

# Create clustered bar plot of missclassified results. 
# Code inspired by https://r-graph-gallery.com/48-grouped-barplot-with-ggplot2.
# Tabulate the number of points correctly and incorrectly classified into clusters 1
# and 2 from Euclidean distance.
cluster_table <- table(group1,Euclidean.clusters)
cluster_table
clustering <- c(rep("Euclidean Cluster 1",2), rep("Euclidean Cluster 2",2))
true_clustering <- rep(c("True Cluster 1 (2027)" , "True Cluster 2 (1630)"), 2)
values <- c(cluster_table[1,1],cluster_table[2,1],cluster_table[1,2],cluster_table[2,2])
data <- data.frame(clustering,true_clustering,values)

ggplot(data, aes(fill=true_clustering, y=values, x=clustering)) + 
  geom_bar(position="dodge", stat="identity") +
  xlab("Clustering") +
  ylab("Number of Points per cluster") +
  geom_col(position = "dodge") +
  geom_text(
    aes(label = values),
    colour = "black", size = 4,
    vjust = -0.3, position = position_dodge(.9)
  ) +
  labs(fill=" ") 

##########################################################################
##########################################################################

# Use the K-Means algorithm with Manhattan distance to classify the points.
set.seed(4511)
res.Manhattan <- Kmeans(df,2,iter.max=100,nstart=100,method="manhattan")
Manhattan.clusters <- res.Manhattan$cluster
Manhattan.centers <- res.Manhattan$centers

# Cluster plot from Manhattan distance.
fviz_cluster(object=list(data=df,cluster=Manhattan.clusters),palette=colours2,repel=TRUE,
             geom="point",show.clust.cent=TRUE,ellipse=TRUE,ellipse.type="convex")

# Create clustered bar plot of missclassified results. 
# Code inspired by https://r-graph-gallery.com/48-grouped-barplot-with-ggplot2.
cluster_table <- table(group1,Manhattan.clusters)
cluster_table
clustering <- c(rep("Manhattan Cluster 1",2), rep("Manhattan Cluster 2",2))
true_clustering <- rep(c("True Cluster 1 (2027)" , "True Cluster 2 (1630)"), 2)
values <- c(cluster_table[1,1],cluster_table[2,1],cluster_table[1,2],cluster_table[2,2])
data <- data.frame(clustering,true_clustering,values)

ggplot(data, aes(fill=true_clustering, y=values, x=clustering)) + 
  geom_bar(position="dodge", stat="identity") +
  xlab("Clustering") +
  ylab("Number of Points per cluster") +
  geom_col(position = "dodge") +
  geom_text(
    aes(label = values),
    colour = "black", size = 4,
    vjust = -0.3, position = position_dodge(.9)
  ) +
  labs(fill=" ") 
##########################################################################
##########################################################################

# Use the K-Means algorithm with Maximum distance to classify the points.
set.seed(4511)
res.max <- Kmeans(df,2,iter.max=100,nstart=100,method="maximum")
max.clusters <- res.max$cluster
max.centers <- res.max$centers

# Cluster plot from Maximum distance.
fviz_cluster(object=list(data=df,cluster=max.clusters),palette=colours2,repel=TRUE,
             geom="point",show.clust.cent=TRUE,ellipse=TRUE,ellipse.type="convex")

# Create clustered bar plot of missclassified results. 
# Code inspired by https://r-graph-gallery.com/48-grouped-barplot-with-ggplot2.
cluster_table <- table(group1,max.clusters)
cluster_table
clustering <- c(rep("Maximum Cluster 1",2), rep("Maximum Cluster 2",2))
true_clustering <- rep(c("True Cluster 1 (2027)" , "True Cluster 2 (1630)"), 2)
values <- c(cluster_table[1,1],cluster_table[2,1],cluster_table[1,2],cluster_table[2,2])
data <- data.frame(clustering,true_clustering,values)

ggplot(data, aes(fill=true_clustering, y=values, x=clustering)) + 
  geom_bar(position="dodge", stat="identity") +
  xlab("Clustering") +
  ylab("Number of Points per cluster") +
  geom_col(position = "dodge") +
  geom_text(
    aes(label = values),
    colour = "black", size = 4,
    vjust = -0.3, position = position_dodge(.9)
  ) +
  labs(fill=" ") 

##########################################################################
##########################################################################

# Use the K-Means algorithm with Mahalanobis distance to classify the points.
# Do 50 iterations with Euclidean distance first to find initial clustering.
set.seed(4511)
res.Mahalanobis <- Kmeans(df,2,iter.max=50,nstart=50,method="euclidean")
Mahalanobis.clusters <- res.Mahalanobis$cluster
Mahalanobis.centers <- res.Mahalanobis$centers
cluster1_inds <- which(Mahalanobis.clusters==1)
cluster2_inds <- which(Mahalanobis.clusters==2)
centroid1 <- Mahalanobis.centers[1,]
centroid2 <- Mahalanobis.centers[2,]
cov1 <- cov(df[cluster1_inds,])
cov2 <- cov(df[cluster2_inds,])

# Cluster plot of starting point for the Mahalanobis distance.
fviz_cluster(object=list(data=df,cluster=Mahalanobis.clusters),palette=colours2,
             repel=TRUE,geom="point",show.clust.cent=TRUE,ellipse=TRUE,
             ellipse.type="convex")

# Create clustered bar plot of missclassified results for starting point of Mahalanobis
# distance. Code inspired by https://r-graph-gallery.com/48-grouped-barplot-with-ggplot2.
cluster_table <- table(group1,Mahalanobis.clusters)
cluster_table
clustering <- c(rep("Euclidean Cluster 1",2), rep("Euclidean Cluster 2",2))
true_clustering <- rep(c("True Cluster 1 (2027)" , "True Cluster 2 (1630)"), 2)
values <- c(cluster_table[1,1],cluster_table[2,1],cluster_table[1,2],cluster_table[2,2])
data <- data.frame(clustering,true_clustering,values)

ggplot(data, aes(fill=true_clustering, y=values, x=clustering)) + 
  geom_bar(position="dodge", stat="identity") +
  xlab("Clustering") +
  ylab("Number of Points per cluster") +
  geom_col(position = "dodge") +
  geom_text(
    aes(label = values),
    colour = "black", size = 4,
    vjust = -0.3, position = position_dodge(.9)
  ) +
  labs(fill=" ") 

# Perform the Mahalanobis distance part of the algorithm.
for (i in 1:100){
  Mahalanobis.clusters <- ifelse(
    mahalanobis(df,centroid1,cov1) < mahalanobis(df,centroid2,cov2), 
    1,
    2
  )
  cluster1_inds <- which(Mahalanobis.clusters==1)
  cluster2_inds <- which(Mahalanobis.clusters==2)
  centroid1 <- colMeans(df[cluster1_inds,])
  centroid2 <- colMeans(df[cluster2_inds,])
  cov1 <- cov(df[cluster1_inds,])
  cov2 <- cov(df[cluster2_inds,])
}

# Cluster plot from Mahalanobis distance.
fviz_cluster(object=list(data=df,cluster=Mahalanobis.clusters),palette=colours2,
             repel=TRUE,geom="point",show.clust.cent=TRUE,ellipse=TRUE,
             ellipse.type="convex")

# Create clustered bar plot of missclassified results. 
# Code inspired by https://r-graph-gallery.com/48-grouped-barplot-with-ggplot2.
cluster_table <- table(group1,Mahalanobis.clusters)
cluster_table
clustering <- c(rep("Mahalanobis Cluster 1",2), rep("Mahalanobis Cluster 2",2))
true_clustering <- rep(c("True Cluster 1 (2027)" , "True Cluster 2 (1630)"), 2)
values <- c(cluster_table[1,1],cluster_table[2,1],cluster_table[1,2],cluster_table[2,2])
data <- data.frame(clustering,true_clustering,values)

ggplot(data, aes(fill=true_clustering, y=values, x=clustering)) + 
  geom_bar(position="dodge", stat="identity") +
  xlab("Clustering") +
  ylab("Number of Points per cluster") +
  geom_col(position = "dodge") +
  geom_text(
    aes(label = values),
    colour = "black", size = 4,
    vjust = -0.3, position = position_dodge(.9)
  ) +
  labs(fill=" ")

##########################################################################
##########################################################################

# Decide on the number of clusters.
fviz_nbclust(df, kmeans, method = "wss") +
  geom_vline(xintercept = 2, linetype = 4)

##########################################################################
##########################################################################

# Choose two classes to keep for the second investigation.
inds2 <- which(dataset[,7]=="SEKER"|dataset[,7]=="SIRA")
data2 <- dataset[inds2,1:6]
group2 <- dataset[inds2,7]
group2 <- ifelse(group2=="SIRA",1,2)
table(group2)

# Scale the data.
df <- scale(data2)

# View the appearance of the cluster plot given the correct clustering.
fviz_cluster(object=list(data=df,cluster=group2),palette=colours,repel=TRUE,
             geom="point",show.clust.cent=TRUE,ellipse=TRUE,ellipse.type="convex")

##########################################################################
##########################################################################

# Use the K-Means algorithm with Euclidean distance to classify the points.
set.seed(1234)
res.Euclid <- Kmeans(df,2,iter.max=100,nstart=100,method="euclidean")
Euclidean.clusters <- res.Euclid$cluster
Euclidean.centers <- res.Euclid$centers

# Cluster plot from Euclidean distance.
fviz_cluster(object=list(data=df,cluster=Euclidean.clusters),palette=colours2,repel=TRUE,
             geom="point",show.clust.cent=TRUE,ellipse=TRUE,ellipse.type="convex")

# Create clustered bar plot of missclassified results. 
# Code inspired by https://r-graph-gallery.com/48-grouped-barplot-with-ggplot2.
# Tabulate the number of points correctly and incorrectly classified into clusters 1
# and 2 from Euclidean distance.
cluster_table <- table(group2,Euclidean.clusters)
cluster_table
clustering <- c(rep("Euclidean Cluster 1",2), rep("Euclidean Cluster 2",2))
true_clustering <- rep(c("True Cluster 1 (2636)" , "True Cluster 2 (2027)"), 2)
values <- c(cluster_table[1,1],cluster_table[2,1],cluster_table[1,2],cluster_table[2,2])
data <- data.frame(clustering,true_clustering,values)

ggplot(data, aes(fill=true_clustering, y=values, x=clustering)) + 
  geom_bar(position="dodge", stat="identity") +
  xlab("Clustering") +
  ylab("Number of Points per cluster") +
  geom_col(position = "dodge") +
  geom_text(
    aes(label = values),
    colour = "black", size = 4,
    vjust = -0.3, position = position_dodge(.9)
  ) +
  labs(fill=" ") 

##########################################################################
##########################################################################

# Use the K-Means algorithm with Manhattan distance to classify the points.
set.seed(1234)
res.Manhattan <- Kmeans(df,2,iter.max=100,nstart=100,method="manhattan")
Manhattan.clusters <- res.Manhattan$cluster
Manhattan.centers <- res.Manhattan$centers

# Cluster plot from Manhattan distance.
fviz_cluster(object=list(data=df,cluster=Manhattan.clusters),palette=colours2,repel=TRUE,
             geom="point",show.clust.cent=TRUE,ellipse=TRUE,ellipse.type="convex")

# Create clustered bar plot of missclassified results. 
# Code inspired by https://r-graph-gallery.com/48-grouped-barplot-with-ggplot2.
cluster_table <- table(group2,Manhattan.clusters)
cluster_table
clustering <- c(rep("Manhattan Cluster 1",2), rep("Manhattan Cluster 2",2))
true_clustering <- rep(c("True Cluster 1 (2636)" , "True Cluster 2 (2027)"), 2)
values <- c(cluster_table[1,1],cluster_table[2,1],cluster_table[1,2],cluster_table[2,2])
data <- data.frame(clustering,true_clustering,values)

ggplot(data, aes(fill=true_clustering, y=values, x=clustering)) + 
  geom_bar(position="dodge", stat="identity") +
  xlab("Clustering") +
  ylab("Number of Points per cluster") +
  geom_col(position = "dodge") +
  geom_text(
    aes(label = values),
    colour = "black", size = 4,
    vjust = -0.3, position = position_dodge(.9)
  ) +
  labs(fill=" ") 
##########################################################################
##########################################################################

# Use the K-Means algorithm with Maximum distance to classify the points.
set.seed(1234)
res.max <- Kmeans(df,2,iter.max=100,nstart=100,method="maximum")
max.clusters <- res.max$cluster
max.centers <- res.max$centers

# Cluster plot from Maximum distance.
fviz_cluster(object=list(data=df,cluster=max.clusters),palette=colours2,repel=TRUE,
             geom="point",show.clust.cent=TRUE,ellipse=TRUE,ellipse.type="convex")

# Create clustered bar plot of missclassified results. 
# Code inspired by https://r-graph-gallery.com/48-grouped-barplot-with-ggplot2.
cluster_table <- table(group2,max.clusters)
cluster_table
clustering <- c(rep("Maximum Cluster 1",2), rep("Maximum Cluster 2",2))
true_clustering <- rep(c("True Cluster 1 (2636)" , "True Cluster 2 (2027)"), 2)
values <- c(cluster_table[1,1],cluster_table[2,1],cluster_table[1,2],cluster_table[2,2])
data <- data.frame(clustering,true_clustering,values)

ggplot(data, aes(fill=true_clustering, y=values, x=clustering)) + 
  geom_bar(position="dodge", stat="identity") +
  xlab("Clustering") +
  ylab("Number of Points per cluster") +
  geom_col(position = "dodge") +
  geom_text(
    aes(label = values),
    colour = "black", size = 4,
    vjust = -0.3, position = position_dodge(.9)
  ) +
  labs(fill=" ") 

##########################################################################
##########################################################################

# Use the K-Means algorithm with Mahalanobis distance to classify the points.
# Do 50 iterations with Euclidean distance first to find initial clustering.
set.seed(1234)
res.Mahalanobis <- Kmeans(df,2,iter.max=50,nstart=50,method="euclidean")
Mahalanobis.clusters <- res.Mahalanobis$cluster
Mahalanobis.centers <- res.Mahalanobis$centers
cluster1_inds <- which(Mahalanobis.clusters==1)
cluster2_inds <- which(Mahalanobis.clusters==2)
centroid1 <- Mahalanobis.centers[1,]
centroid2 <- Mahalanobis.centers[2,]
cov1 <- cov(df[cluster1_inds,])
cov2 <- cov(df[cluster2_inds,])

# Cluster plot of starting point for the Mahalanobis distance.
fviz_cluster(object=list(data=df,cluster=Mahalanobis.clusters),palette=colours2,
             repel=TRUE,geom="point",show.clust.cent=TRUE,ellipse=TRUE,
             ellipse.type="convex")

# Create clustered bar plot of missclassified results for starting point of Mahalanobis
# distance. Code inspired by https://r-graph-gallery.com/48-grouped-barplot-with-ggplot2.
cluster_table <- table(group2,Mahalanobis.clusters)
cluster_table
clustering <- c(rep("Euclidean Cluster 1",2), rep("Euclidean Cluster 2",2))
true_clustering <- rep(c("True Cluster 1 (2636)" , "True Cluster 2 (2027)"), 2)
values <- c(cluster_table[1,1],cluster_table[2,1],cluster_table[1,2],cluster_table[2,2])
data <- data.frame(clustering,true_clustering,values)

ggplot(data, aes(fill=true_clustering, y=values, x=clustering)) + 
  geom_bar(position="dodge", stat="identity") +
  xlab("Clustering") +
  ylab("Number of Points per cluster") +
  geom_col(position = "dodge") +
  geom_text(
    aes(label = values),
    colour = "black", size = 4,
    vjust = -0.3, position = position_dodge(.9)
  ) +
  labs(fill=" ") 

# Perform the Mahalanobis distance part of the algorithm.
for (i in 1:100){
  Mahalanobis.clusters <- ifelse(
    mahalanobis(df,centroid1,cov1) < mahalanobis(df,centroid2,cov2), 
    1,
    2
  )
  cluster1_inds <- which(Mahalanobis.clusters==1)
  cluster2_inds <- which(Mahalanobis.clusters==2)
  centroid1 <- colMeans(df[cluster1_inds,])
  centroid2 <- colMeans(df[cluster2_inds,])
  cov1 <- cov(df[cluster1_inds,])
  cov2 <- cov(df[cluster2_inds,])
}

# Cluster plot from Mahalanobis distance.
fviz_cluster(object=list(data=df,cluster=Mahalanobis.clusters),palette=colours2,
             repel=TRUE,geom="point",show.clust.cent=TRUE,ellipse=TRUE,
             ellipse.type="convex")

# Create clustered bar plot of missclassified results. 
# Code inspired by https://r-graph-gallery.com/48-grouped-barplot-with-ggplot2.
cluster_table <- table(group2,Mahalanobis.clusters)
cluster_table
clustering <- c(rep("Mahalanobis Cluster 1",2), rep("Mahalanobis Cluster 2",2))
true_clustering <- rep(c("True Cluster 1 (2636)" , "True Cluster 2 (2027)"), 2)
values <- c(cluster_table[1,1],cluster_table[2,1],cluster_table[1,2],cluster_table[2,2])
data <- data.frame(clustering,true_clustering,values)

ggplot(data, aes(fill=true_clustering, y=values, x=clustering)) + 
  geom_bar(position="dodge", stat="identity") +
  xlab("Clustering") +
  ylab("Number of Points per cluster") +
  geom_col(position = "dodge") +
  geom_text(
    aes(label = values),
    colour = "black", size = 4,
    vjust = -0.3, position = position_dodge(.9)
  ) +
  labs(fill=" ")

##########################################################################
##########################################################################

# Decide on the number of clusters.
fviz_nbclust(df, kmeans, method = "wss") +
  geom_vline(xintercept = 2, linetype = 4)
\end{lstlisting}

\end{document}